\documentclass[review]{elsarticle}

\usepackage{lineno,hyperref,fullpage}
\usepackage{afterpage}

\usepackage{enumitem}

\journal{Elsevier}

\usepackage{nomencl}
\usepackage{amsmath}
\usepackage{amsfonts}
\usepackage{multirow}
\usepackage{subcaption}

\usepackage{floatrow}
\makenomenclature
\usepackage{xpatch}
\xpatchcmd{\thenomenclature}{\section*{\nomname}
}{}{\typeout{Success}}{\typeout{Failure}}
\makenomenclature
\RequirePackage{ifthen}
\printnomenclature[0.8cm]

\usepackage{soul}
\usepackage{xcolor}
\usepackage{bm}

\usepackage{url}

\hypersetup{nolinks=true}
\renewcommand\nomgroup[1]{%
  \item[\itshape
  \ifstrequal{#1}{A}{Symbols}{%
  \ifstrequal{#1}{B}{Roman Letters}{%
  \ifstrequal{#1}{C}{Greek Letters}{%
  \ifstrequal{#1}{D}{Abbreviations}{}}}}%
]}

\let\oldequation\equation
\let\oldendequation\endequation
\renewenvironment{equation}
  {\linenomathNonumbers\oldequation}
  {\oldendequation\endlinenomath}

\begin{document}

\begin{frontmatter}

    \title{Physical interpretation of neural network-based nonlinear eddy viscosity models}

  \author[cas,ucas]{Xin-Lei Zhang}

  \author[sim]{Heng Xiao}

  \author[gist]{Solkeun Jee \corref{mycor}}
  \cortext[mycor]{Corresponding author}
  \ead{sjee@gist.ac.kr}
  
  \author[cas,ucas]{Guowei He}

  \address[cas]{The State Key Laboratory of Nonlinear Mechanics, Institute of Mechanics, Chinese Academy of Sciences, Beijing 100190, China}
  \address[ucas]{School of Engineering Sciences, University of Chinese Academy of Sciences, Beijing 100049, China}
  \address[sim]{Stuttgart Center for Simulation Science (SC SimTech), University of Stuttgart, Stuttgart, Germany}
  \address[gist]{School of Mechanical Engineering, Gwangju Institute of Science and Technology, Gwangju 61005, South Korea}

\begin{abstract}
Neural network-based turbulence modeling has gained significant success in improving turbulence predictions by incorporating high--fidelity data.
However, the interpretability of the learned model is often not fully analyzed, which has been one of the main criticism of neural network-based turbulence modeling.
Therefore, it is increasingly demanding to provide physical interpretation of the trained model, which is of significant interest for guiding the development of interpretable and unified turbulence models.
The present work aims to interpret the predictive improvement of turbulence flows based on the behavior of the learned model, represented with tensor basis neural networks.
The ensemble Kalman method is used for model learning from sparse observation data due to its ease of implementation and high training efficiency.
Two cases, i.e., flow over the S809 airfoil and flow in a square duct, are used to demonstrate the physical interpretation of the ensemble-based turbulence modeling.
For the flow over the S809 airfoil, our results show that the ensemble Kalman method learns an optimal linear eddy viscosity model, which improves the prediction of the aerodynamic lift by reducing the eddy viscosity in the upstream boundary layer and promoting the early onset of flow separation.
For the square duct case, the method provides a nonlinear eddy viscosity model, which predicts well secondary flows by capturing the imbalance of the Reynolds normal stresses.
The flexibility of the ensemble-based method is highlighted to capture characteristics of the flow separation and secondary flow by adjusting the nonlinearity of the turbulence model.
\end{abstract}

  \begin{keyword}
     Machine learning \sep turbulence modeling \sep ensemble Kalman inversion \sep physical interpretability
  \end{keyword}
\end{frontmatter}

\section{Introduction}

Data-driven turbulence modeling has emerged as an important approach for predicting turbulent flows~\cite{duraisamy2019turbulence}, which constructs functional mappings from mean velocity to the Reynolds stress by incorporating observation data.
Over the past few years, this paradigm of turbulence modeling has been pursued  from various aspects, including choice of training data, development of training strategy, and representative form of the Reynolds stress.
As for the training data, both the Reynolds stress and velocity data have been used for learning turbulence models.
The velocity data become advocated for model learning as they are relatively straightforward to obtain in practical applications compared to the Reynolds stress data~\cite{strofer2021end}.
Regarding the training strategies, the conventional \textit{a priori} approach~\cite{ling2015evaluation,ling2016reynolds, zhu2019machine} trains neural network-based models without involving the RANS solver, which is pointed out~\cite{duraisamy2021perspectives} to have inconsistency issues in posterior tests.
For this reason, the model--consistent training~
\cite{holland2019field,zhao2020rans,macart2021embedded,strofer2021end,zhang_ensemble-based_2022,wang2023unified} has been proposed to improve the predictive abilities of learned models by coupling the neural network and the RANS equation during the training process.
Besides the two research lines mentioned above, the representative form of turbulence closure has also been investigated to empower the model with generalizability across different classes of flows.
It is one critical step toward the ultimate goal of discovering unified turbulence models from data.

Various strategies have been proposed to represent the Reynolds stress, such as neural-network-based multiplicative correction~\cite{singh2017machine}, eigen perturbation method~\cite{wang2017physics,wu2018physics,wu2019representation}, symbolic expression~\cite{weatheritt2016novel}, tensor basis neural network~\cite{ling2016reynolds} and so on.
Specifically, the neural-network-based multiplicative correction is introduced to modify turbulent production terms in turbulence transport equations, which can improve the velocity prediction of separated flows but is still under the Bounssinesq assumption.
The eigen perturbation method is proposed to present the Reynolds stress based on the eigen decomposition of the Reynolds stress tensor.
The obtained eigenfunctions have physical interpretations to indicate the magnitude, shape, and orientation of the Reynolds stress tensor.
This representation is a general form to represent the Reynolds stress but requires careful selection of the input features to ensure the Galilean invariance.
To overcome these limitations,  the nonlinear eddy viscosity model is often used as the base model, which is beyond the Boussinesq assumption and regards scalar invariants associated with velocity gradients as model inputs.
Different techniques, including symbolic expression and neural networks, have been introduced to represent the Reynolds stress based on the nonlinear eddy viscosity model.
In this work, we focus on the neural network-based representation, i.e., tensor basis neural network~\cite{ling2016reynolds}.

The tensor basis neural network is able to represent the anisotropy of the Reynolds stress flexibly due to its great expressive power.
Neural networks have expressive power that increases exponentially with the depth of the network. 
Hence, it has the potential to achieve a universal or at least unified model to represent various flow characteristics.
That is, one model form is applicable to multiple classes of flows, such as attached flows, separated flows, and corner flows, possibly with internal switching or branching.
While such universality is not the objective of this work, it is appealing to have such possibilities in the future.
However, the tensor basis neural network has intrinsic drawbacks due to the weak equilibrium assumption and the black-box feature.
On the one hand, although the tensor basis neural network is the most general nonlinear eddy viscosity model, it is still a local model under the weak equilibrium assumption.
That is, the Reynolds stress anisotropy only depends on the local velocity gradient.
To address this issue, the vector--cloud neural network~\cite{zhou2022frame} has been proposed to enforce the nonlocal dependence in the representative form.
On the other hand, the trained neural network is still a black box and has encountered an interpretability crisis for neural network-based turbulence modeling.
Therefore, it is of significant necessity to interpret the physical mechanism behind the learned neural network and guide the development of turbulence closures.

In this work, we aim to physically interpret the behavior of the learned turbulence model in terms of predictive improvement.
Neural networks can represent complex functional relationships between physical quantities but have poor interpretability on the learned model behavior.
In contrast, symbolic models are often assumed as interpretable since they can provide the causes and effects of the model behavior in the \textit{a priori} sense.
It is noted that when the learned symbolic model provides a complicated expression that is highly composited or has many high-order terms, which would also be difficult to interpret. 
Some post--hoc approaches, such as the Shapley additive explanations (SHAP) method~\cite{Lundberg2017shap}, have been proposed to interpret the black-box neural network models.
These methods can indicate the importance value of each input feature on the neural network output~\cite{he2022explainability} in the \textit{a posteriori} sense.
However, they cannot provide physical insights into the mechanism of the learned model for improving the RANS prediction.

In this work, we investigate the physical interpretability of the learned turbulence model, represented with tensor basis neural networks~\cite{ling2016reynolds}.
The ensemble Kalman method is adopted to learn turbulence models from sparse observation data, including the lift force and velocity.
We show that the behavior of the learned neural network is physically interpretable to improve flow predictions on two canonical flows, i.e., separated flow in the S809 airfoil and secondary flow in a square duct.
The ensemble method can adjust the nonlinearity of the learned model to capture the different flow characteristics.
Moreover, the capability of the ensemble Kalman method is shown in learning turbulence models from very sparse observation data.
In addition, the normalization strategy is investigated to avoid feature clustering due to the stagnation point of airfoil flows.
We note that the interpretability in this work refers to the model behavior of the trained neural network.
It is different from the interpretability of neural networks in the machine learning community, which aims to present the features of neural networks in an understandable term, e.g., indicating the importance of input features with specific contribution values~\cite{Lundberg2017shap}.

The rest of the paper is outlined as follows.
The ensemble-based modeling methodology is elaborated in Section~\ref{sec:method}.
The case setups and the training results are presented in Section~\ref{sec:case_setup} and~\ref{sec:results}, respectively.
Finally, the paper is concluded in Section~\ref{sec:conclusion}.

\section{Methodology}  
\label{sec:method}

For incompressible turbulent flows, the mean flow can be described by the RANS equation as
\begin{equation}
\begin{aligned}
    \nabla \cdot \boldsymbol{u} &= 0  \\
    \boldsymbol{u} \cdot \nabla \boldsymbol{u}  &= - \nabla p + \nu \nabla^2 \boldsymbol{u} - \nabla \cdot \boldsymbol{\tau} \text{,}
\end{aligned}
\end{equation}
where $p$ is the mean pressure normalized by the flow density, $\boldsymbol{u}$ is the velocity vector, $\nu$ represents the molecular viscosity, and $\boldsymbol{\tau}$
indicates the Reynolds stress\footnote{Here we followed Pope's convention~\cite{pope2001turbulent} of defining Reynolds stress as the covariance of the velocity fluctuations i.e., $\tau_{ij} = \left<u^{'}_i u^{'}_j \right>$. We note that in the literature (e.g.,~\cite{wilcox2006turbulence}) it is more common to call $- \left< u^{'}_i u^{'}_j \right>$ the Reynolds stress because of its role in the RANS momentum equations.}  to be modeled.
Here we aim to construct neural-network-based turbulence models by incorporating available observations, such as lift force and velocity measurements.
In the following, we introduce the Reynolds stress representation and the ensemble-based training method adopted in this work.

\subsection{Neural-network-based turbulence closure}

The tensor basis neural network~\cite{ling2016reynolds} is used to represent the Reynolds stress due to the flexibility to represent the anisotropy of Reynolds stress.
In the tensor basis neural network, the Reynolds stress~$\boldsymbol{\tau}$ is decomposed into a deviatoric part and an isotropic part, as
\begin{equation}
\begin{aligned}
    \boldsymbol{\tau} &= 2k \sum_{\ell=1}^{10} g^{(\ell)} \mathbf{T}^{(\ell)} + \frac{2k}{3} \mathbf{I} \text{,}  \\ 
    \text{with} \quad g^{(\ell)} &= g^{(\ell)}\left(\theta_{1}, \ldots, \theta_{5}\right) \text{,}
\end{aligned}
\label{eq:tau}
\end{equation}
where $k$ is the turbulent kinetic energy, $\mathbf{T}$ is the tensor basis, $g^{(\ell)}$ is the coefficient of the tensor basis to be determined, $\boldsymbol{\theta}$ is the scalar invariants, and $\mathbf{I}$ is the identity matrix.
The $g$ functions are represented with neural networks in this work which approximates functional mappings from the scalar invariants~$\boldsymbol{\theta}$ to the basis coefficients.
There are ten independent tensor bases based on the Cayley-Hamilton theory~\cite{pope1975more} and five scalar invariants for incompressible flows.
In the 2D scenario, only two scalar invariants and three tensor bases are remained~\cite{pope2001turbulent}.
Further, the third tensor basis can be incorporated in the pressure term for incompressible flows, leaving only two scalar invariants.
The first four tensor bases can be written as
\begin{equation}
\begin{aligned}
    \mathbf{T}^{(1)} &= \hat{\mathbf{S}}, \qquad \mathbf{T}^{(2)} = \hat{\mathbf{S}} \hat{\mathbf{W}} - \hat{\mathbf{W}}\hat{\mathbf{S}}, \\ 
    \mathbf{T}^{(3)} &= \hat{\mathbf{S}}^2 - \frac{1}{3}\{\hat{\mathbf{S}}^2\} \mathbf{I}, \qquad \mathbf{T}^{(4)} = \hat{\mathbf{W}}^2-\frac{1}{3}\{\hat{\mathbf{W}}^2\}\mathbf{I} \text{.}
    \label{eq:tensor_basis}
\end{aligned}
\end{equation}
In the formula above, $\{\cdot\}$ denotes the trace operator, and the $\hat{\mathbf{S}}$ and $\hat{\mathbf{W}}$ are the normalized strain rate and the rotation rate based on the turbulence time scale~$\tau_s$, i.e., 
\begin{equation}
\begin{aligned}
    \hat{\mathbf{S}} &= \tau_s \mathbf{S} \quad \hat{\mathbf{W}} = \tau_s \mathbf{W} \\
    \text{with} \quad \mathbf{S} &= \frac{1}{2}(\nabla \boldsymbol{u} + \nabla \boldsymbol{u}^\top) \quad 
    \text{and} \quad \mathbf{W} = \frac{1}{2} (\nabla \boldsymbol{u} - \nabla  \boldsymbol{u}^\top) \text{.}
\end{aligned}
\end{equation}
The time scale~$\tau_s$ can be estimated with the turbulent kinetic energy~$k$ and the dissipation rate~$\varepsilon$ or the specific dissipation rate~$\omega$.
It is noted that the time scale becomes zero as we approach the wall. Hence one can bound the time scale with the Kolmogorov scale~\cite{durbin1993application} as
\begin{equation}
    \tau_s = \text{max}\left(\frac{k}{\varepsilon}, C_\tau \sqrt{\frac{\nu}{\varepsilon}}\right) \text{,}
\end{equation}
where $C_\tau$ is constant and set as $6$ in this work.

\subsection{Normalization of input features}
The input features of the neural networks should be scaled within $[-1,1]$ to accelerate the training convergence.
The min--max normalization is able to confine the input features within the range of $[0, 1]$ (see e.g., Ref.~\cite{strofer2021end}).
A normalized feature~$\hat{\theta}$ can be formulated as
$\hat{\theta} = (\theta - \theta_\text{min})/(\theta_\text{max} - \theta_\text{min})$, 
where the subscript `min' and `max' indicate the minimum and maximum value of a given feature~$\theta$.  
However, when there exist singular points with extremely large magnitudes in computational domains, this normalization strategy can lead to severe feature clustering.
For instance, the velocity gradient near a stagnation point can have an extremely large value.
Using the global maximum value to normalize entire input features will lead to most feature values clustering around $0$, which would significantly affect the training performance.

In this work, the scalar invariants~$\boldsymbol{\hat{\theta}}$ are normalized  with the local time scale~$\tau_s$~\cite[e.g.,][]{ling2015evaluation,wu2019representation,liu2023learning} based on 
\begin{equation}
\begin{aligned}
    \hat{\theta}_1 &= \{\tilde{\mathbf{S}}^2 \}, \qquad \hat{\theta}_2 = \{\tilde{\mathbf{W}}^2\}, \\ 
    \tilde{\mathbf{S}} &= \frac{\mathbf{S}}{\| \mathbf{S} \| + 1 / \tau_s}, \quad \text{and} \qquad 
    \tilde{\mathbf{W}} = \frac{\mathbf{W}}{\| \mathbf{W} \|+ 1 / \tau_s} \text{.}
    \label{eq:scalar_invariant}
\end{aligned}
\end{equation}
With this specific normalization, the scalar invariants can be scaled within $[-1, 1]$ to avoid feature clustering along certain directions.
The normalized scalar invariants~$\boldsymbol{\theta}$ are used as the neural network inputs, and the coefficients~$g$ of the tensor bases are regarded as the outputs.
Further, the neural network outputs~$g$ are combined with the tensor bases~$\mathbf{T}$ to form the anisotropic part of the Reynolds stress.
The obtained Reynolds stress is used to predict the velocity and pressure fields by solving the RANS equations.
Moreover, the constructed Reynolds stress~$\boldsymbol{\tau}$ is used to compute the turbulence production term in the turbulent kinetic energy and dissipation rate transport equations.
Further, the neural network weights are optimized by incorporating observation data based on the ensemble Kalman method, which will be illustrated in the following subsection.

\subsection{Model-consistent training with ensemble Kalman method}

Model--consistent training couples a neural network and a CFD solver during the training process.
By doing this, it can ensure consistency between the training and prediction environments, thereby alleviating the ill-conditioning of the RANS model operator~\cite{wu2019reynolds}.
Moreover, this strategy can leverage sparse observation data, e.g., velocity measurements, to train the neural network-based model.
This is in contrast to the prior training where the model is often trained with the full field data of the Reynolds stress and has poor generalizability due to the inconsistency issue~\cite{zhang_ensemble-based_2022}.
The model-consistent training amounts to finding the optimal weights of neural networks that lead to the best fit with the sparse observation data. 

Various training methods can be used to perform the model--consistent training, including the adjoint method~\cite{holland2019field}, the ensemble method~\cite{zhang_ensemble-based_2022}, and the genetic programming method~\cite{zhao2020rans}.
We use the ensemble Kalman method for model training due to its non-derivative nature and good training efficiency.
The ensemble method is a statistical inference method that uses an ensemble of samples to guide the optimization~\cite{strofer2021ensemble}, which has been used for the physical modeling of subsurface flows~\cite{zhou2023inference} and turbulent flows with high Reynolds numbers~\cite{wang2023unified,liu2023learning}.
We use this method to train the turbulence model represented with the tensor basis neural network.
The update scheme of the ensemble Kalman method 
can be formulated as
\begin{equation}
\begin{aligned}
    & \boldsymbol{w}_j^{i+1} = \boldsymbol{w}_j^i + \mathsf{K}(\mathsf{y}_j-\mathsf{H}\boldsymbol{w}_j^i) \\
    & \text{with} \quad \mathsf{K} = \mathsf{PH}^\top(\mathsf{HPH}^\top + \mathsf{R})^{-1} \text{.}
    \label{eq:enkf}
\end{aligned}
\end{equation}
Herein $\mathsf{H}$ is the local gradient of the model prediction~$\mathcal{H}[\boldsymbol{w}]$ with respect to the weights of neural networks $\boldsymbol{w}$, $\mathsf{P}$ is the model error covariance, $\mathsf{R}$ is the observation error covariance, $\mathsf{y}$ is the observation data, and $i$ and $j$ represent the index of optimization iteration and sample, respectively.
The model operator~$\mathsf{H}$ is often avoided to be computed by reformulating the Kalman gain matrix as
$$ \mathsf{K} = \mathsf{S}_w \mathsf{S}_y^\top(\mathsf{S}_y \mathsf{S}_y^\top + \mathsf{R})^{-1} \text{.} $$
The square-root matrices $\mathsf{S}_w$ and $\mathsf{S}_y$ are defined as
\begin{subequations}
    \label{eq:sqrt_root}
    \begin{align}
    \mathsf{S}_w^i &= \dfrac{1}{\sqrt{N_e -1}} \left[\bm{w}_1^i - \overline{\bm{w}}^i, \bm{w}_2^i - \overline{\bm{w}}^i, \dotsb, \bm{w}_{N_e}^i - \overline{\bm{w}}^i\right], \\
    \mathsf{S}_y^i &= \dfrac{1}{\sqrt{N_e -1}} \left[\mathcal{H}[\bm{w}_1^i] - \mathcal{H}[\overline{\bm{w}}^i], \mathcal{H}[\bm{w}_2^i] - \mathcal{H}[\overline{\bm{w}}^i], \dotsb, \mathcal{H}[\bm{w}_{N_e}^i] - \mathcal{H}[\overline{\bm{w}}^i]\right] , \label{eq:sqrt_root_y} \\
    \overline{\bm{w}}^i&= \dfrac{1}{N_e} \sum_{j=1}^{N_e} \bm{w}_j^i \text{,}
    \end{align}
\end{subequations}
which are estimated from the samples at every iteration.
In this work, we use the ensemble-based Kalman update scheme for learning turbulence models in a model-consistent manner.
As pointed out in Ref.~\cite{zhang_ensemble-based_2022}, in scenarios having large data sets, e.g., time-dependent three-dimensional flow fields, the present algorithm would be computationally expensive and need to incorporate reduced-order techniques such as the truncated singular value decomposition~\cite{luo2018correlation}.

Note that the ensemble Kalman method can train the neural network-based model with multiple observation data, including the measurements at various flow conditions.
Specifically, we can incorporate the observation data at different flow conditions sequentially.
It is achieved by training neural networks with each observation data in several inner loops. 
The maximum iteration number of the inner loop is set as $3$ in this work based on our sensitivity study.
Moreover, the observation data are shuffled randomly before training, which allows escaping from local minima similar to the stochastic gradient descent method~\cite{bottou2003stochastic}.
Further, the Kalman update scheme is used to incorporate the observation data in the shuffled order till the entire data sets are traversed.
After that, the observation data will be reshuffled and continue to be incorporated with the ensemble Kalman method.
The practical implementation of the ensemble method is presented in~\ref{sec:implementation}.
One can also augment the observation with data from different flow conditions.
However, this may drop into local minima and lead to unsatisfactory predictive accuracy in certain cases since the ensemble method aims to reduce the L2 norm of the total data misfit.
By shuffling the training data, the method is able to find the global minimum and provide more accurate turbulence models based on our numerical tests.

\section{Case setup}
\label{sec:case_setup}

We use two cases to demonstrate the physical interpretation of the ensemble-based turbulence modeling, i.e., the flow over the S809 airfoil and the flow in a square duct.
The two cases represent canonical separated flows and secondary flows, respectively.
Both are challenging for conventional linear eddy viscosity models.
The distinct flow characteristics are able to examine the flexibility of the ensemble-based method in learning interpretable models from partial observation.
The details of the case setup are described in the following subsections.

\subsection{Flow over S809 airfoil}

Flow over the S809 airfoil has been widely used for numerical validation of turbulence models as well as their data-driven counterparts~\cite{singh2017machine,holland2019field}.
Such flows are challenging for linear eddy viscosity models at large angles of attack due to the flow separation.
Conventional RANS models cannot accurately predict the massive flow separation, which further leads to the overestimation of the lift force beyond the stall angle~\cite{singh2016using}.
Here we aim to interpret the behavior of the neural network-based model learned from lift force measurements with the ensemble Kalman method.

\begin{figure}[!htb]
    \centering
    \subfloat[Computational domain]{\includegraphics[height=0.4\textwidth]{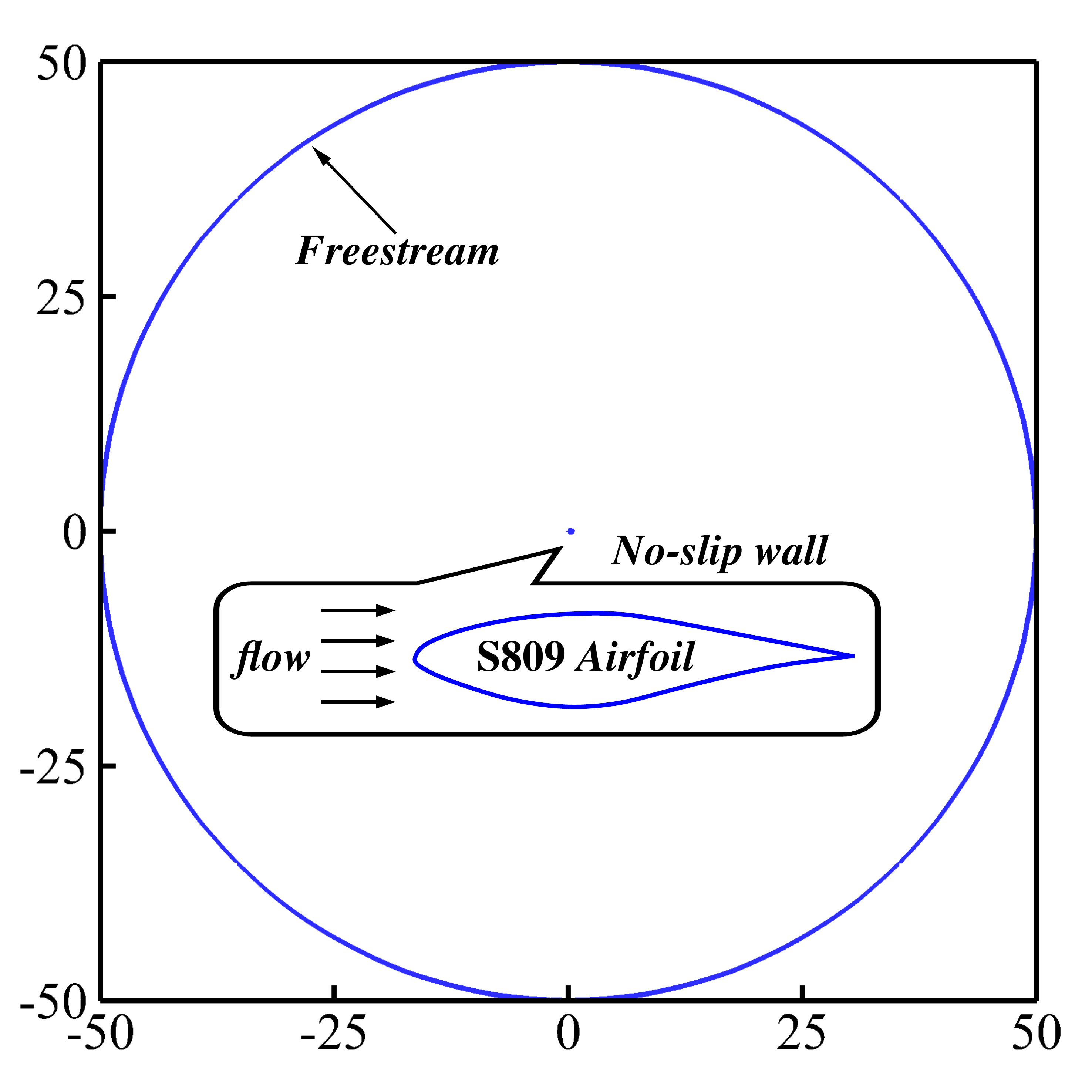}}
    \subfloat[Mesh grid]{\includegraphics[height=0.4\textwidth]{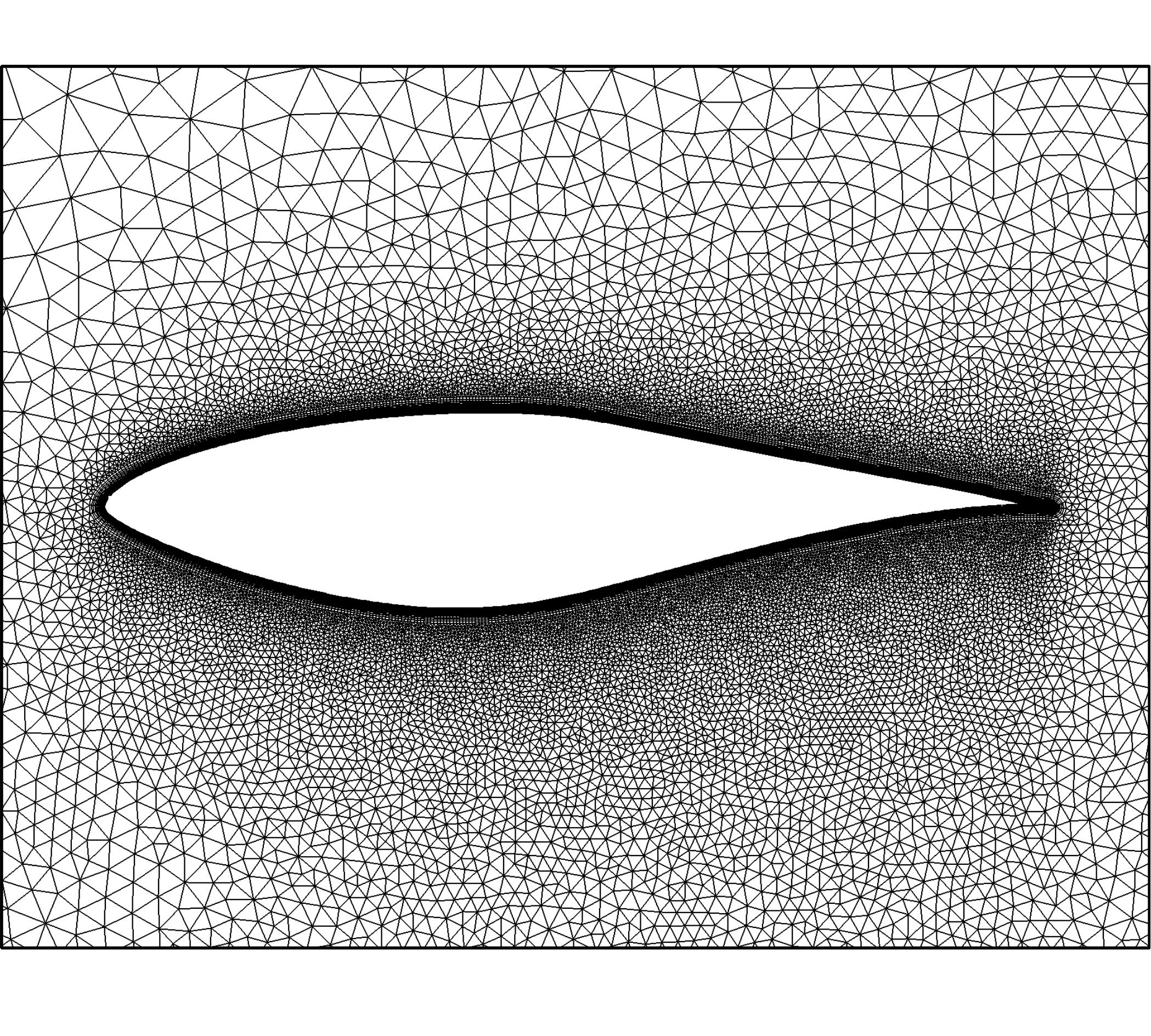}}
    \caption{ Computational domain and mesh grid for computations of flows over the S809 airfoil}
    \label{fig:mesh_S809}
\end{figure}

The Reynolds number is $Re_c=2 \times 10^6$ based on the inflow velocity and chord length.
The angle of attack~$\alpha$ varies from $1^\circ$--$18^\circ$.
At large angles of attack, conventional turbulence models underestimate the separation zones~\cite{singh2016using}, which leads to large discrepancies in the predictions of the lift force~\cite{singh2016using}.
The unstructured mesh with around $78000$ cells is used to discretize the computational domain.
The mesh grid from the work~\cite{holland2019field} is adopted in the work as shown in Fig.~\ref{fig:mesh_S809}.
The no-slip condition is employed on the airfoil surface.
The height of the first cell in the normal direction corresponds to $y^+ \approx 1$.

For the flow around the S809 airfoil, the available measurement data is the lift force, which is the integral type data source.
Such limited observation would increase the ill-posedness of the inverse problems~\cite{zhang2020regularized,zhang2022assessment,zhang2023combining}.
That is, different model functions can provide similar lift forces.
Further, the learned model could have poor predictive accuracy and robustness due to the ill-posedness issue.
To alleviate the issue, we use observations at two angles of attack, i.e., $8.2^\circ$ and $14.24^\circ$ to train the model.
The former is attached flows, while the latter is separated flows.
Learning from both the attached and separated flows can provide a model with better predictive ability.

As for the setup of the ensemble-based learning algorithm, the number of samples is taken as $50$.
The initial relative variance of the samples is set as $0.1$, which is used to draw the random samples.
The measurements of the lift force~\cite{osti_437668} are used as training data.
The relative observation error is set as $0.01$.
The first two scalar invariants~$\theta_{1}$ and $\theta_2$, and the first two tensor coefficients $g^{(1)}$ and $g^{(2)}$ are used as the inputs and the outputs of the neural network, respectively.
We use the $k$--$\omega$ model~\cite{wilcox2006turbulence} as the baseline model, which is suitable for complex boundary layer flows with adverse pressure gradient compared to standard $k$--$\varepsilon$ model~\cite{launder1974application}.

\subsection{Flow in a square duct}

The secondary flow in a square duct is mainly driven by the imbalance of the Reynolds normal stresses~$\tau_{yy}-\tau_{zz}$~\cite{speziale_turbulent_1982}.
The linear eddy viscosity model is not able to predict the secondary flow since it cannot well estimate the anisotropy of the Reynolds stress.
We use this case to demonstrate the flexibility of the ensemble method in building interpretable nonlinear models from sparse observation data of secondary flows.

The Reynolds number based on the bulk velocity and half of the duct height is $Re_h=3500$ for this case.
Only one quadrant of the physical domain is simulated, considering the symmetry of the flow to the centerlines along $y-$ and $z-$axes.
The mesh with $50 \times 50$ is used to discretize the domain.
The non-slip condition is imposed on the wall, and the symmetry condition is imposed at the symmetry boundary.

\begin{figure}
    \centering
    \includegraphics[height=0.36\textwidth]{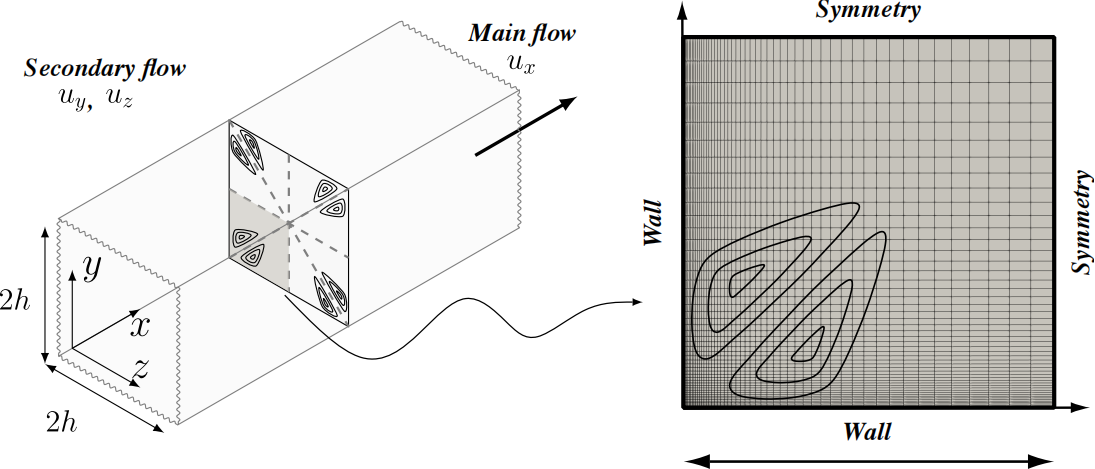}
    \caption{Computational domain and mesh grid for the fully--developed square duct case.}
    \label{fig:mesh_squareDuct}
\end{figure}

As for the setup of the ensemble-based learning algorithm, the number of samples is set as $50$.
The initial variance of the weights is set as $0.1$.
The DNS data~\cite{pinelli_uhlmann_sekimoto_kawahara_2010} are used to train the neural network-based model.
The velocity profiles at $y/H=0.25, 0.5, 0.75, 1.0$ are regarded as the observation data.
The total number of observation data points is $200$.
The relative observation error is set as $0.01$. 
For this case, we use the first two scalar invariants ${\theta}_1$ and $\theta_2$ as the input features, and the first four tensor coefficients~$g^{(1-4)}$ as the model outputs.
Compared to the linear eddy viscosity model, the tensor bases of $\mathbf{T}^{(2)}$, $\mathbf{T}^{(3)}$ and $\mathbf{T}^{(4)}$ are introduced to capture the anisotropy of the Reynolds normal stress, thereby producing the secondary flow.
The $k$--$\varepsilon$ model~\cite{launder1974application} is used as the baseline model in this case, since this model is often taken as the base of nonlinear eddy viscosity model for secondary flows~\cite{shih1993realizable,craft1997prediction}.

    In this work, the open source CFD library OpenFOAM~\cite{opencfd2018openfoam} is used to solve the RANS equations with turbulence models.
    Specifically, the built-in solver \textit{simpleFOAM} is used to solve the RANS equations, given the Reynolds stress fields.
    The Reynolds stresses are constructed with the neural networks, and the scalar invariants from the RANS computation are taken as inputs of the networks.
    The weights of the neural network are updated with the observation data based on the ensemble Kalman method.
    The TensorFlow~\cite{abadi2015tensorflow} library is used to construct the neural network, and the DAFI code~\cite{strofer2021dafi} is used to implement the ensemble Kalman method.
    The test cases and the weights of the learned neural network are publicly available~\cite{zhang2023dafi} for reproducibility.

\begin{table}[!htb]
    \centering
    \begin{tabular}{c|c|c}
        \hline
        Cases & S809 airfoil &   Square duct \\
        \hline
        mesh counts & $\approx 78000$ & 2500 \\
        Reynolds number & $Re_c = 2 \times 10^6$ & $Re_h=3500$ \\
        data & $C_l$ (at $\alpha=8.2^\circ$ and $14.24^\circ$) & $\boldsymbol{u}$ (at $y/H = 0.25, 0.5, 0.75, 1$) \\ 
        baseline model & $k$--$\omega$ & $k$--$\varepsilon$ \\
        initial relative variance & $0.1$ & $0.1$ \\
        relative observation error & $0.01$ & $0.01$ \\
        sample size & $50$ & $50$ \\
        \hline
    \end{tabular}
    \caption{Computational parameters used in the flow around the S809 airfoil and the flow in a square duct}
    \label{tab:test_case}
\end{table}

\section{Results}
\label{sec:results}

\subsection{Flows over S809 airfoil}

\subsubsection{Training performance}

The learned model can improve the predictions of the lift force compared to the baseline $k$--$\omega$ model.
The predicted lift with the learned and baseline models at two chosen angles of attack is listed in Table~\ref{tab:lift}.
The baseline model predicts the aerodynamic lift $C_l=1.25$ at the angle of attack~$14.24^\circ$, which deviates much from the experimental observation~$C_l=1.05$~\cite{osti_437668} due to the massive flow separation.
In contrast, the learned model provides $C_l=1.07$, which is in good agreement with the experimental data.
At $\alpha=8.2^\circ$, the boundary layer is attached, and the baseline model provides good prediction with $C_l=0.97$.
The learned model predicts $C_l=1.0$, which is slightly deviated from the observation $C_l=0.95$.
That is because the learning method decreases the data misfit at the two flow conditions simultaneously.
The significant decrease of lift force at the angle~$\alpha=14.24^\circ$ is achieved with the sacrifice of a slight discrepancy at $\alpha=8.2^\circ$.
In general, the learned model can provide good predictions close to the experimental measurements, which is not surprising since the experimental data are used to train the model function.

\begin{table}[!htb]
    \centering
    \begin{tabular}{c|c|c|c|c}
    \hline
          & $\alpha$ & baseline $k$--$\omega$ & learned model & experiment~\cite{osti_437668}  \\
    \hline
    $C_l$ & $14.24^\circ$    & 1.25 & 1.07 & 1.05 \\
    $C_l$ & $8.2^\circ$      & 0.97 & 1.00 & 0.95 \\
    \hline
    \end{tabular}
    \caption{Summary of the prediction in the aerodynamic lift~$C_l$ with the learned and baseline $k$--$\omega$ models compared to the experimental data for the S809 airfoil}
    \label{tab:lift}
\end{table}

\begin{figure}[!htb]
    \centering
    \subfloat[$\alpha=8.2^\circ$]{\includegraphics[width=0.4\textwidth]{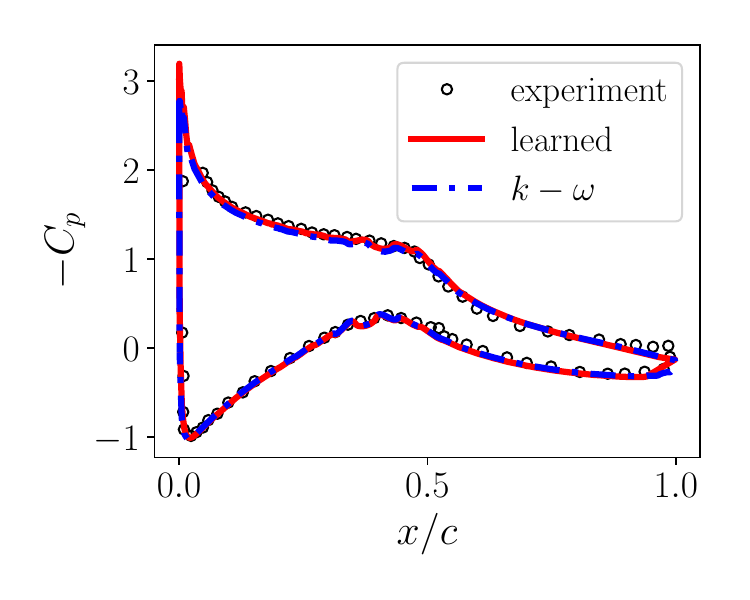}}
    \subfloat[$\alpha=14.24^\circ$]{\includegraphics[width=0.4\textwidth]{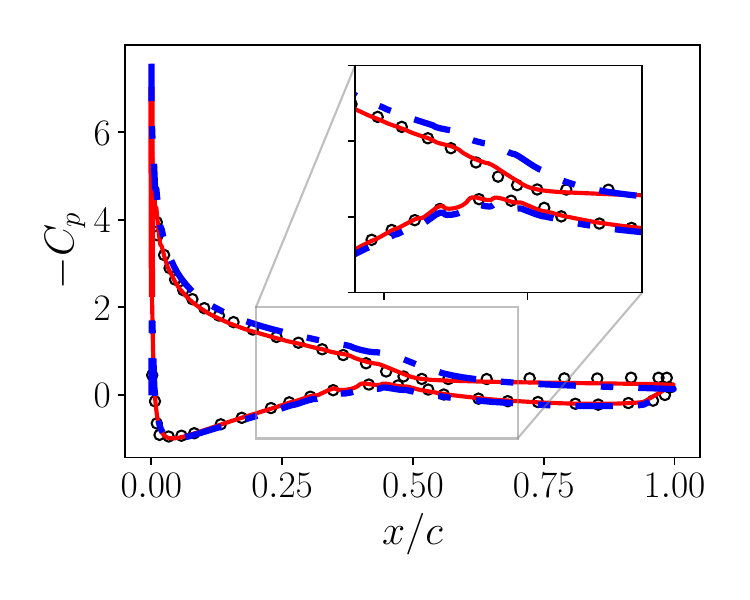}}
    \caption{Wall pressure coefficient~$C_p$ at $\alpha=8.2^\circ$ and $14.24^\circ$ with the learned model and the baseline $k$--$\omega$ model compared to the experimental data~\cite{osti_437668} for the S809 airfoil
    }
    \label{fig:cp}
\end{figure}

The prediction of the pressure coefficient~$C_p$ is improved with the learned model compared to the baseline model.
Figure~\ref{fig:cp} shows the predicted $C_p$ with comparison among the learned model, the baseline $k$--$\omega$ model, and the experimental data.
It can be seen that the baseline $k$--$\omega$ model can predict well the pressure distribution on the surface of the S809 airfoil at~$\alpha=8.2^\circ$.
However, at~$\alpha=14.24^\circ$ the baseline model underestimates the suction pressure on the upper surface of the airfoil.
In contrast, the learned model with the ensemble method is able to predict $C_p$ in better agreement with the experimental data for both angles of attack.
The results demonstrate that the learned model can leverage integral data, i.e., lift force $C_l$, to improve the prediction of wall pressure distribution.

\subsubsection{Physical interpretation of the model behavior}

The learned model can accurately predict the aerodynamic lift and the wall pressure distribution beyond the stall angle compared to the baseline model.
Such improvements can be interpreted by analyzing the model behavior in terms of the friction coefficient, flow separation, and modeled quantities.
Therefore, we further provide comparisons between the learned model and the baseline model in the following.

\begin{figure}
    \centering
    \includegraphics[width=0.9\textwidth]{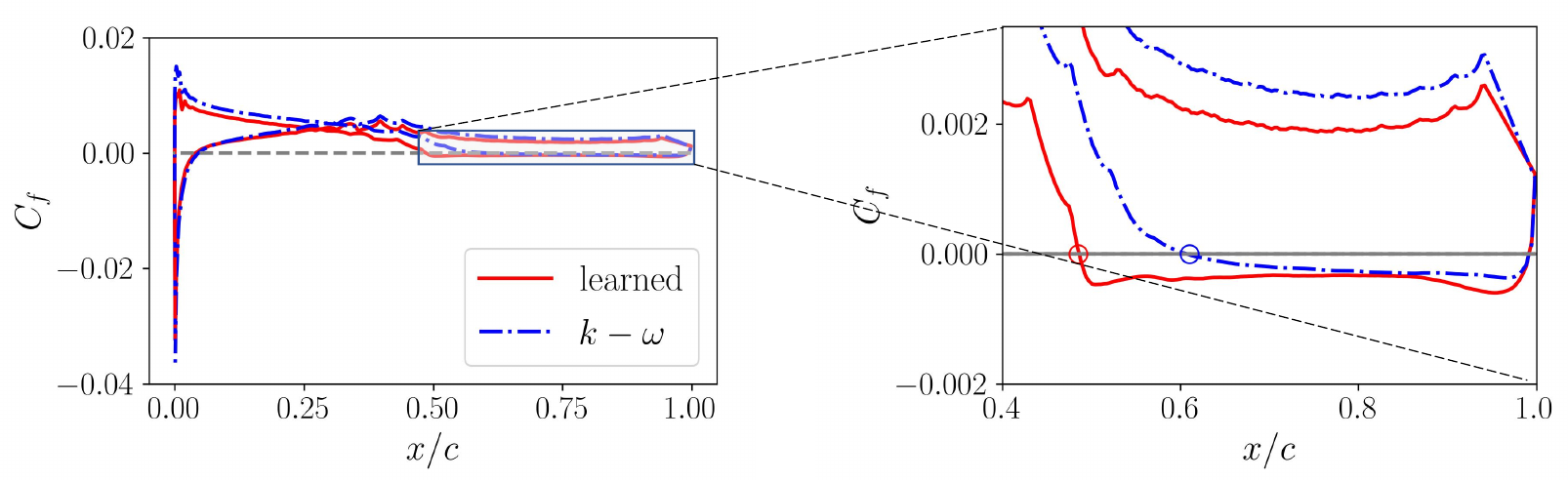}
    \caption{The comparison of the friction coefficient between the learned model and the baseline model at the angle of attack of $14.24^\circ$. The round circles in the right panel indicate the separation locations.}
    \label{fig:cf}
\end{figure}

The friction coefficient~$C_f$ on the airfoil is investigated to interpret the reduction of $C_l$ with the learned model at~$\alpha=14.24^\circ$.
Figure~\ref{fig:cf} shows $C_f$ from the baseline and the learned models.
The learned model leads to the early onset of flow separation, while the baseline model delays the separation.
The early separation would lead to an enlarged re-circulation region.
Therefore, the early occurrence of the flow separation is responsible for the modification of $C_p$ on the suction side as shown in Fig.~\ref{fig:cp}.
Also, it is observed that the friction coefficient is reduced on the pressure side. 
That is because the flow around the airfoil changes globally with the learned model due to the upstream separation point on the upper surface. This global modification of the flow also affects the flow on the lower surface.

\begin{figure}[!htb]
    \centering
    \subfloat[$k$--$\omega$]{\includegraphics[width=0.47\textwidth]{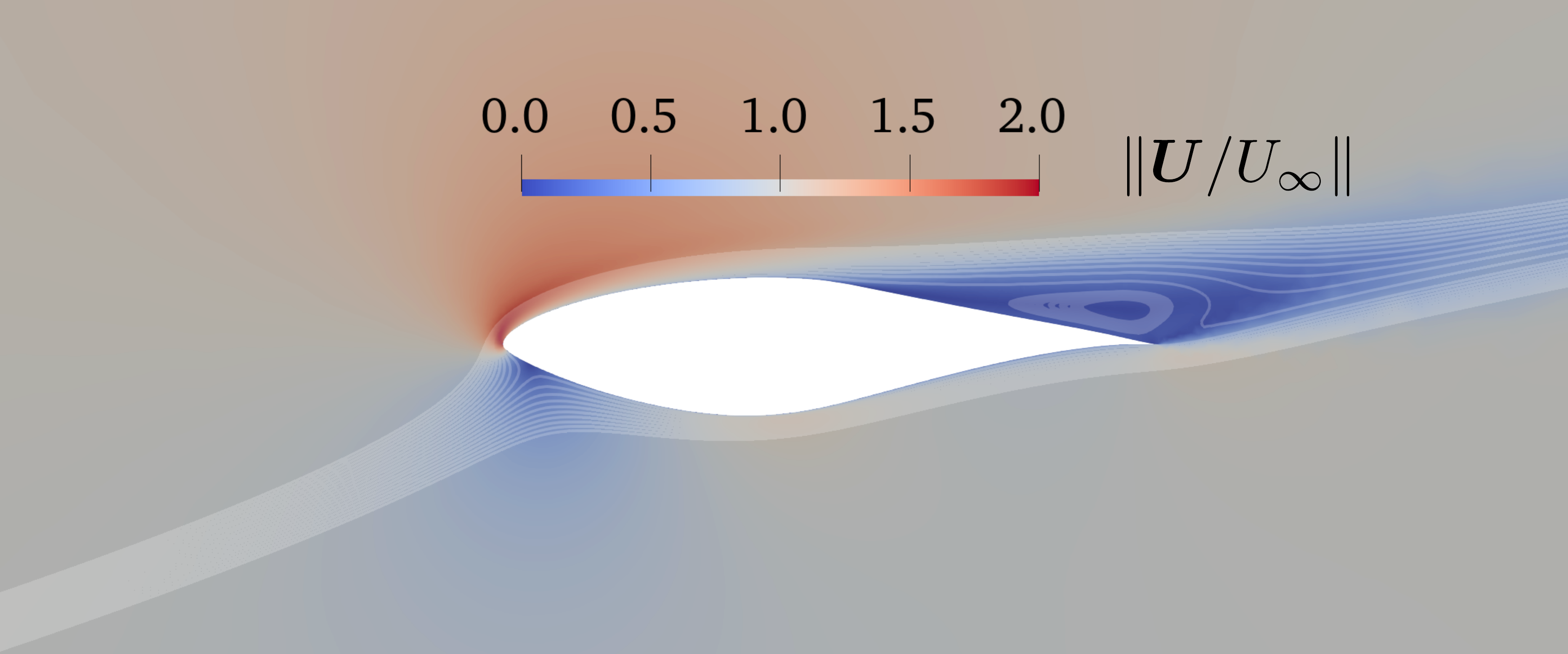}}
    \hfill
    \subfloat[learned model]{\includegraphics[width=0.47\textwidth]{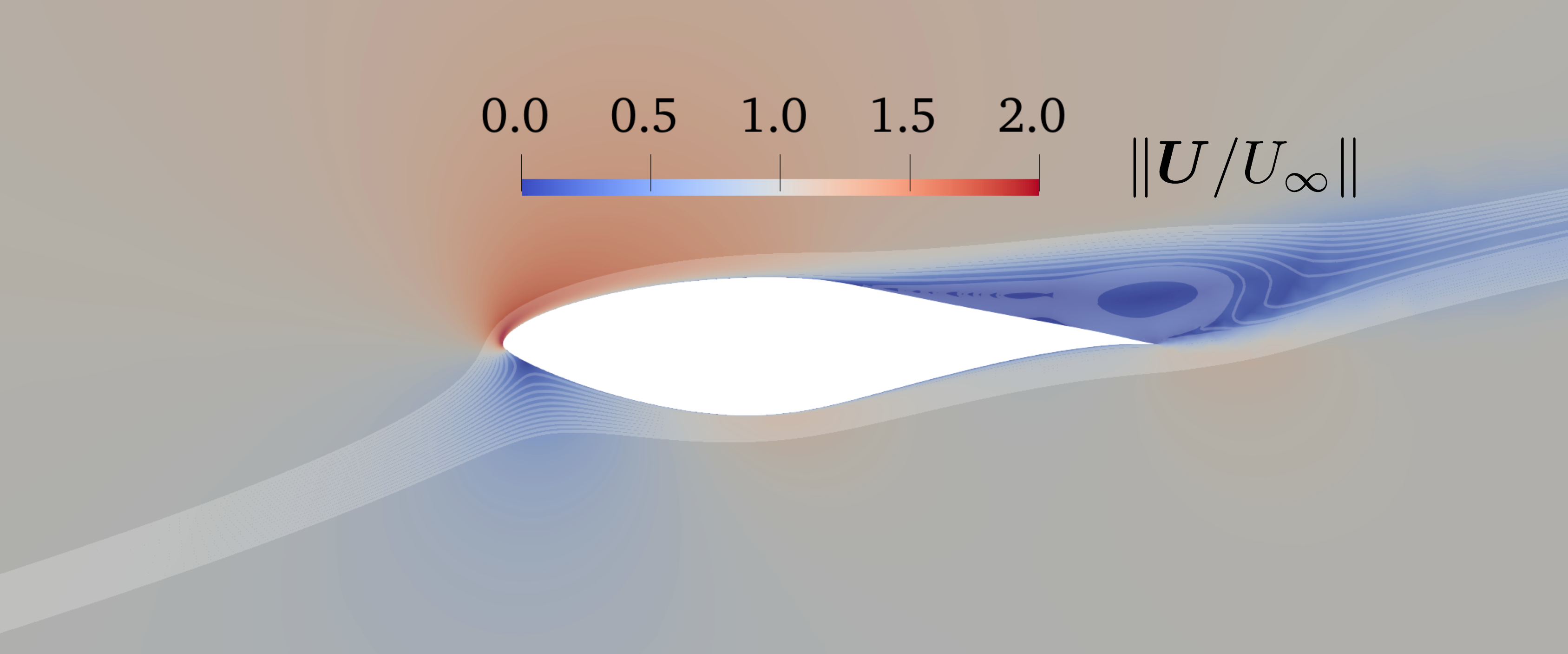}}
    \caption{Predicted separation bubbles at angles of attack~$14.24^\circ$ with the learned model and the $k$--$\omega$ model for the S809 airfoil case.
    }
    \label{fig:bubs}
\end{figure}

The enlarged re-circulation region with the learned model can be clearly seen in Figure~\ref{fig:bubs}, which presents streamlines around the airfoil.
The baseline $k$--$\omega$ model predicts a relatively small separation bubble compared to that with the learned model, which is consistent with the lift prediction listed in Table~\ref{tab:lift}. 
The learned model produces a sufficient massive separation region, which leads to the improvement of $C_l$.

\begin{figure}
    \centering
    \includegraphics[width=0.8\textwidth]{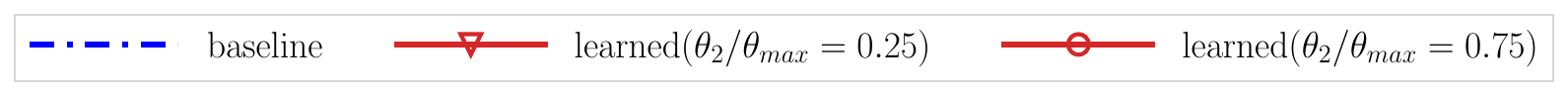}
    \subfloat[$g^{(1)}$ function]{\includegraphics[width=0.45\textwidth]{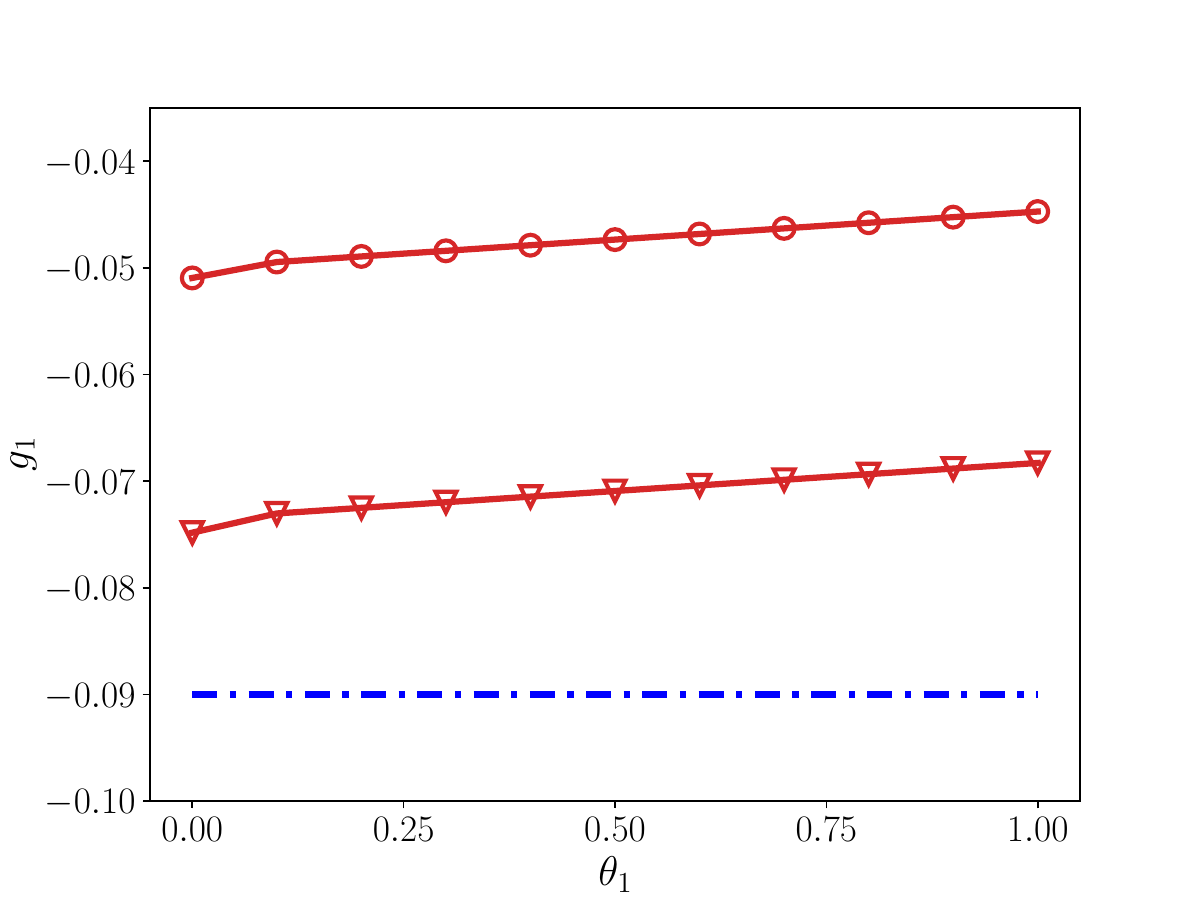}}
    \subfloat[$g^{(2)}$ function]{\includegraphics[width=0.45\textwidth]{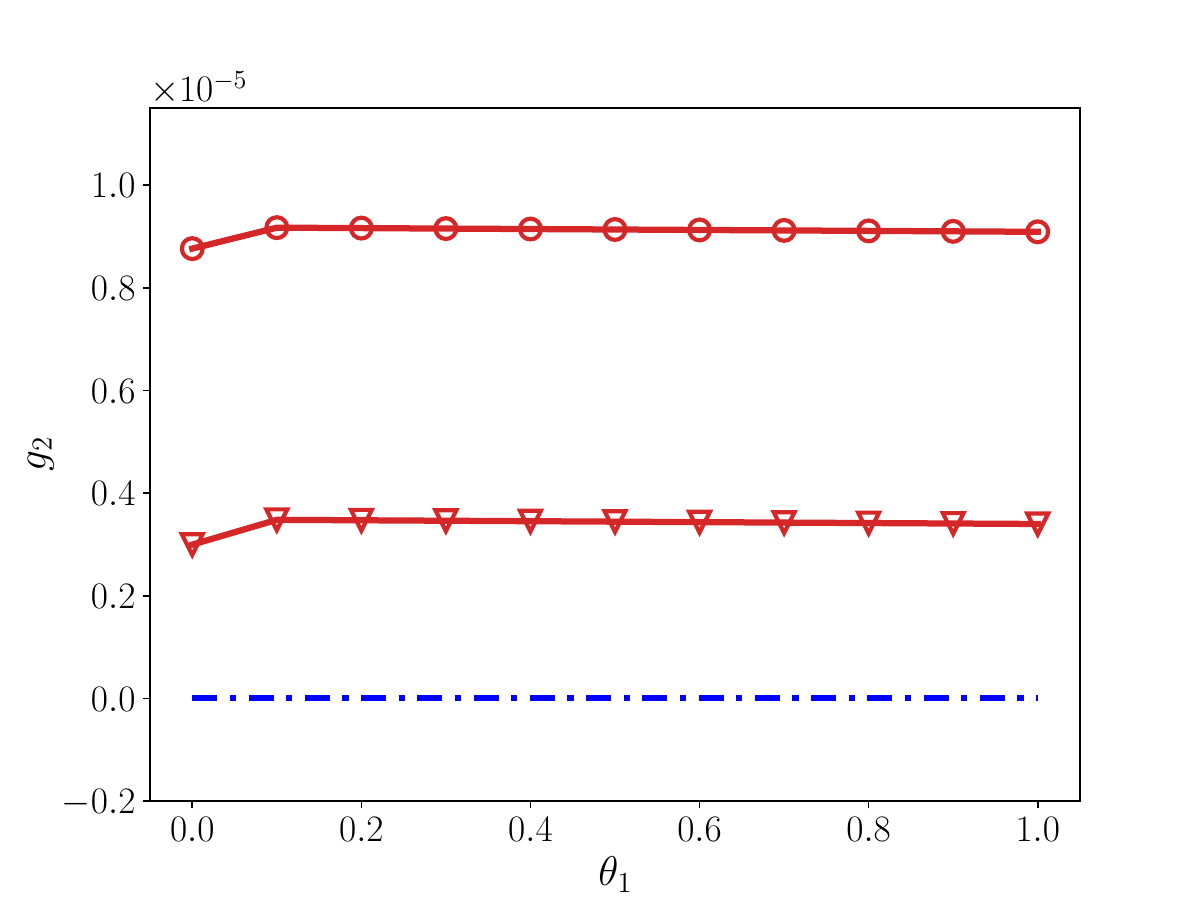}}
    \caption{Plots of the learned mapping between the scalar invariants~$\boldsymbol{\theta}$ and the tensor coefficient~$\boldsymbol{g}$, compared to the baseline for the S809 airfoil case. For the learned model, the plots indicate the learned function at $\theta_2 / \theta_\text{max}=0.25$ and $0.75$.}
    \label{fig:gfunc_s809}
\end{figure}

We investigate the model function~$g$ to interpret the reason for the early onset of the flow separation.
The learned model for the S809 airfoil is almost a linear eddy viscosity model.
It can be seen from Figure~\ref{fig:gfunc_s809} which presents the learned ${g}$ functions on $\theta_1$ at fixed planes of $\theta_2/\theta_\text{max}=0.25$ and $0.75$.
The magnitude of the learned $g^{(1)}$ function decreases to around $0.05$ and $0.075$ at the plane of $\theta_2/\theta_\text{max}=0.25$ and $0.75$, respectively, while the magnitude for the baseline model is constant at $0.09$.
In contrast, the learned $g^{(2)}$ function is almost zero with the order of magnitude of $10^{-6}$, which is similar to the baseline model, i.e., $g^{(2)}=0$.
Therefore, the learned model can be considered a linear eddy viscosity model, which is capable of capturing the flow separation on the S809 airfoil.

\begin{figure}[!htb]
    \centering
     \subfloat[$\| g^{(1)} \mathbf{T^{(1)}} \|$]{\includegraphics[width=0.45\textwidth]{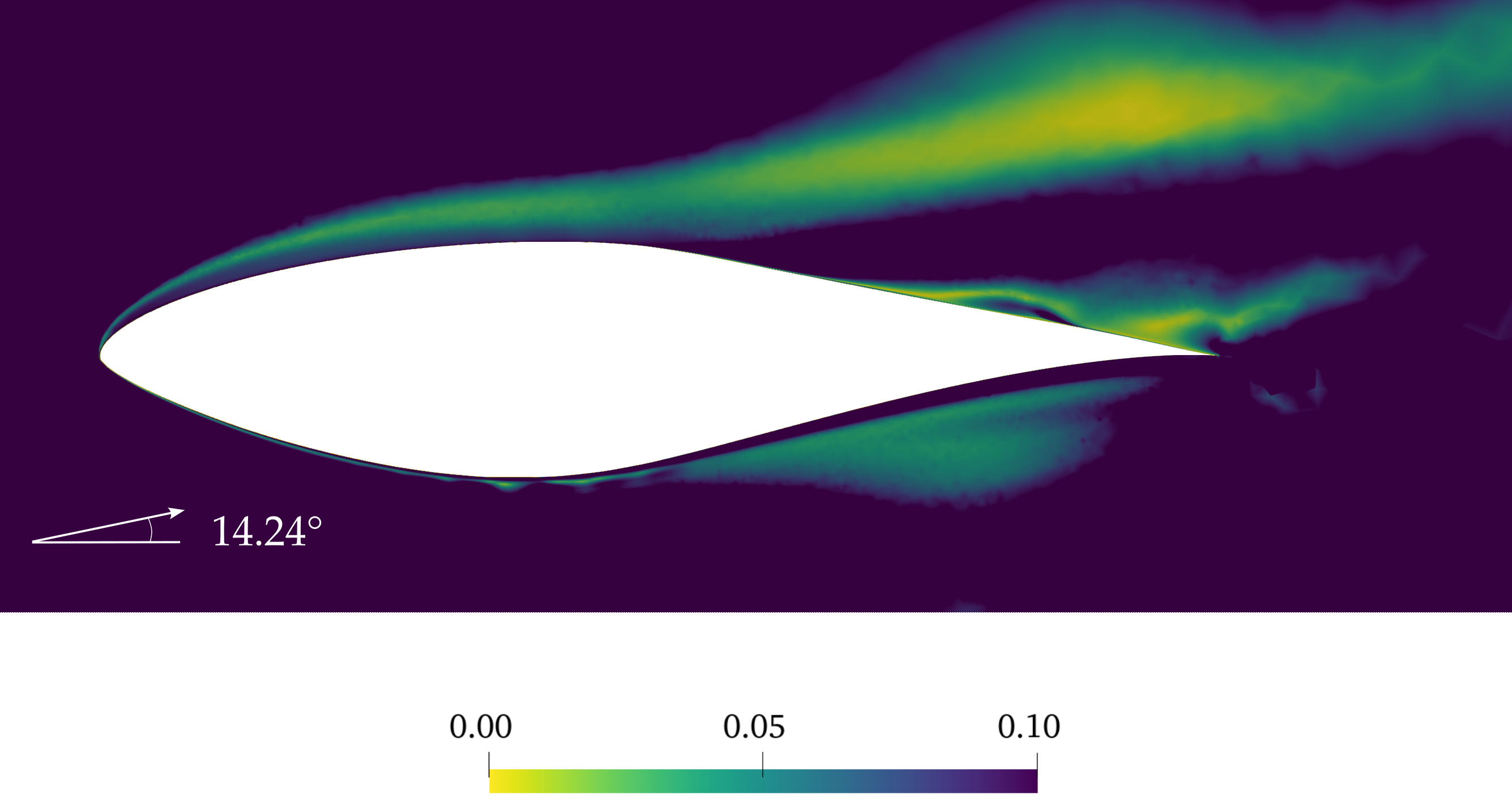}}
     \hfill
    \subfloat[$\| g^{(2)} \mathbf{T^{(2)}} \|$]
    {\includegraphics[width=0.45\textwidth]{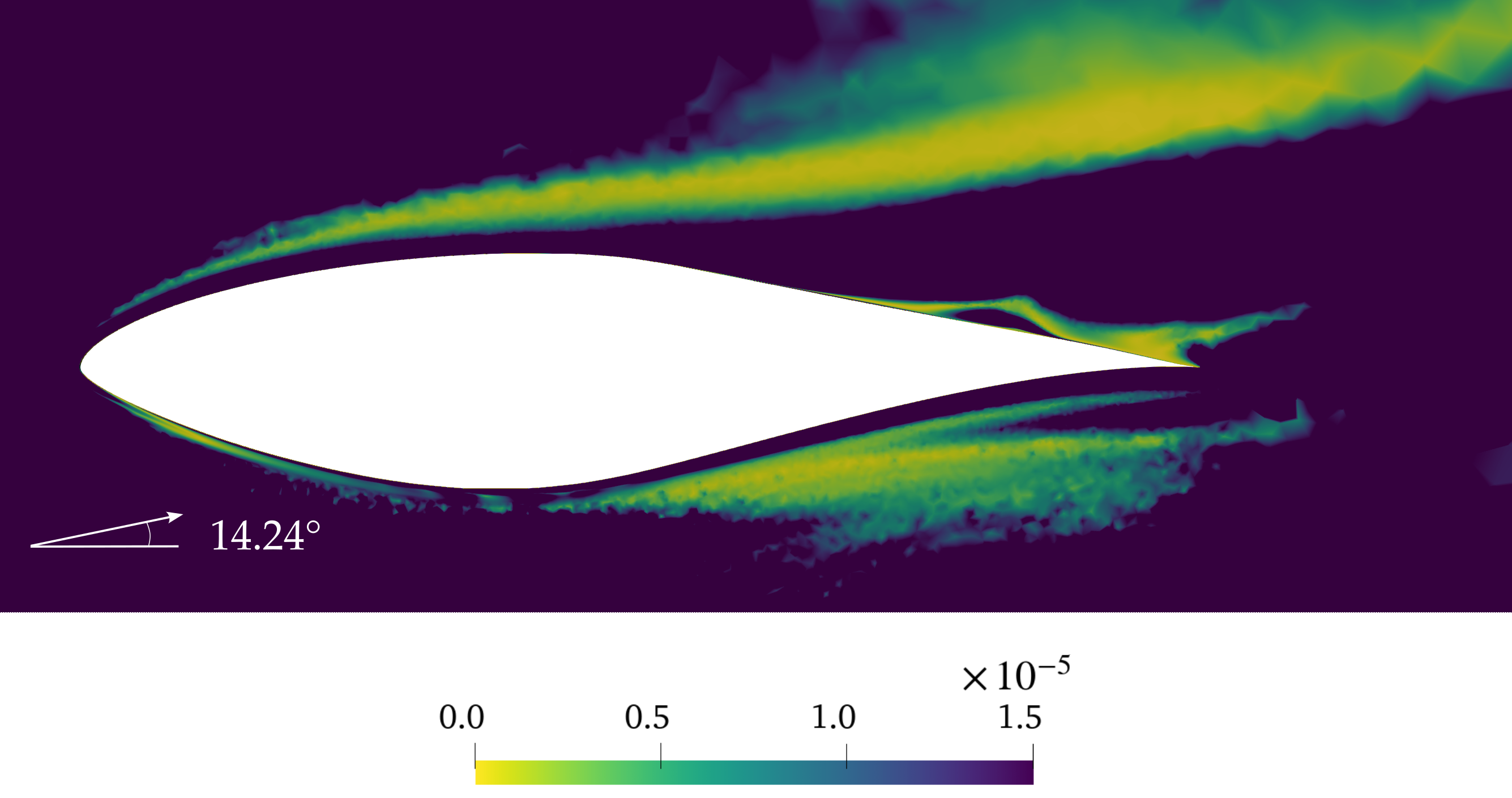}}
    \caption{Learned tensor components for the S809 airfoil case.  The arrow indicates the direction of incoming flow (14.24$^\circ$).}
    \label{fig:tensor_component_s809}
\end{figure}

Recall that the Reynolds stress anisotropy is the linear combination of the learned function~$g(\boldsymbol{\theta})$ and the tensor bases~$\mathbf{T}$, i.e., $\mathbf{b}=\sum g^{(\ell)} \mathbf{T}^{(\ell)}$.
Fig.~\ref{fig:tensor_component_s809} presents the contour plots of the magnitude of each tensor component.
The first term is a linear term on the strain rate (see Eqs.~\eqref{eq:tau} and~\eqref{eq:tensor_basis}).
For the airfoil case, the nonlinear term is almost zero as shown in Fig.~\ref{fig:tensor_component_s809}(b), which further confirms that the learned model can be considered a linear model under the Boussinesq assumption.
Such linear models can also achieve good predictions in the lift force for the 2D airfoil case.
This is consistent with the work of Singh et al.~\cite{singh2016using, singh2017machine}, where a multiplicative correction is added in the turbulence transport equation to modify the eddy viscosity and improve the lift prediction beyond the stall angle.
The linear eddy viscosity assumption is sufficient in this study of the S809 airfoil.

\begin{figure}[!htb]
    \centering
    \subfloat[$k$--$\omega$]{\includegraphics[width=0.48\textwidth]{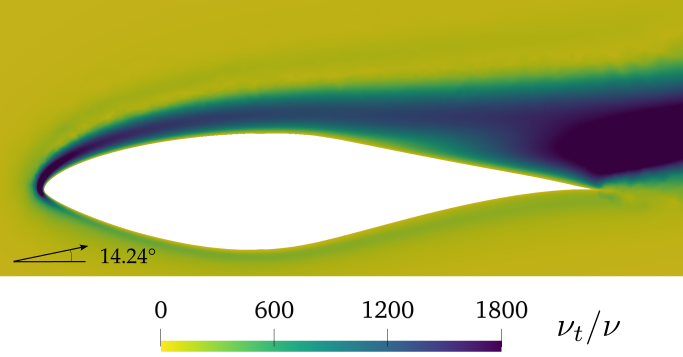}}
    \hfill
    \subfloat[learned model]{\includegraphics[width=0.48\textwidth]{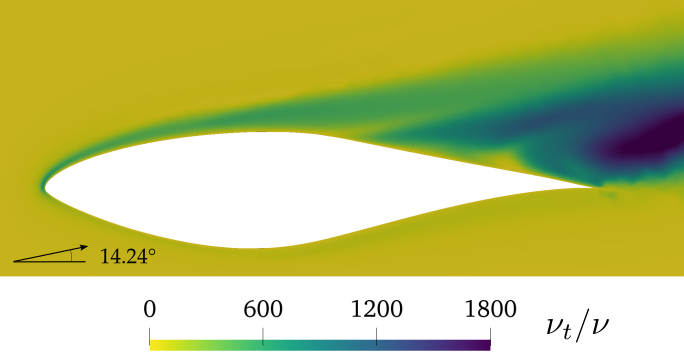}}
    \caption{Comparison of the eddy viscosity between the learned and baseline $k$--$\omega$ models for the S809 airfoil. Note that the eddy viscosity is normalized by the molecular viscosity as $\nu_t / \nu$. The arrow indicates the direction of incoming flow (14.24$^\circ$).
    }
    \label{fig:nut}
\end{figure}

Since the learned model can be regarded as a linear eddy viscosity model, we further investigate the effects of the learned eddy viscosity on the model prediction.
Figure~\ref{fig:nut} shows the predicted eddy viscosity with the learned model compared to the baseline $k$--$\omega$ model.
The eddy viscosity is computed based on the $g^{(1)}$ function, which can be formulated as 
$$\nu_t = -\frac{g^{(1)} k}{C_{\mu} \omega} .$$
The model constant is $C_{\mu} = 0.09$.
It can be seen clearly that the eddy viscosity is reduced, particularly around the upstream boundary layer and the separated region.
The eddy viscosity can transfer the energy from the outer flow to the boundary layer, which is able to restrain momentum reduction and further flow separation.
The reduced eddy viscosity would weaken the energy transfer from the outer flow and enable the boundary layer to be less resistant to the adverse pressure gradient. 
As such, the reduced eddy viscosity in the upstream can induce a relatively early onset of the flow separation and further a large separation region.
We note that the reduced eddy viscosity at the upstream boundary layer is also observed~\cite{Heo2022} from the model learned with the adjoint-based method~\cite{singh2017machine}, which further confirms the physical interpolation for the improved model prediction. 
In addition, the eddy viscosity at the lower surface is reduced as well compared to the baseline $k$--$\omega$ model.
This explains the friction coefficient reduction on the pressure side as presented in Fig.~\ref{fig:cf}, because the wall shear stress (or the velocity gradient) is sensitive to a small change near the wall -- here the eddy viscosity is slightly reduced.

\begin{figure}[!htb]
    \centering
    \includegraphics[width=0.7\textwidth]{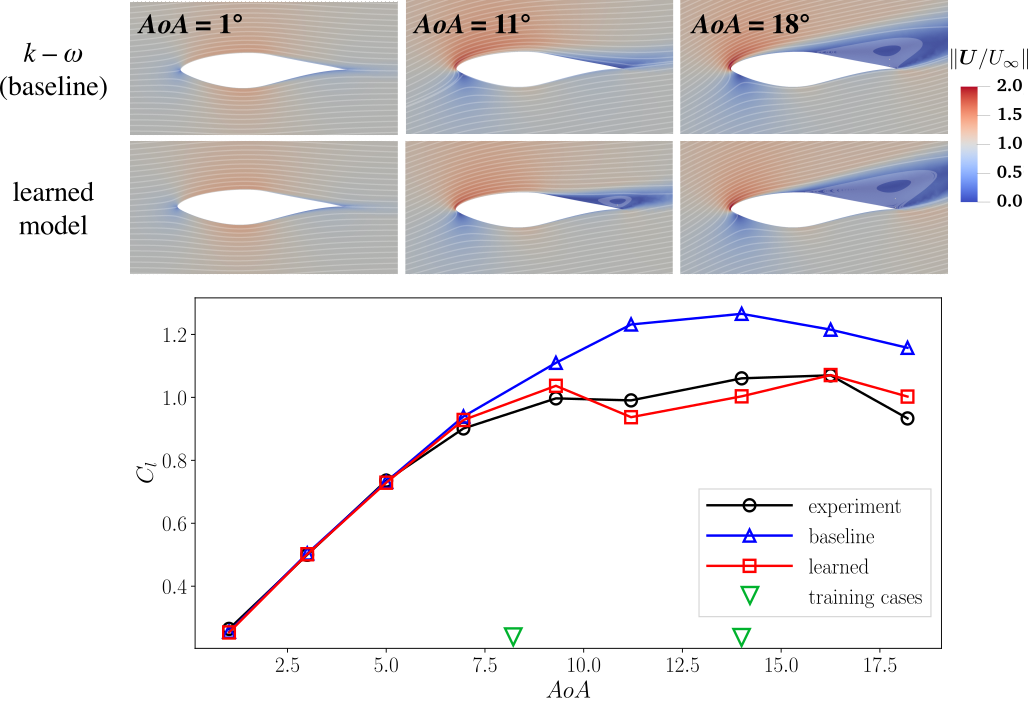}
    \caption{Tests on various angles of attack with comparison among the baseline model, the learned model, and the experiment~\cite{osti_437668} for the S809 airfoil case.
    The training cases are also indicated in the plot.
    }
    \label{fig:generalizability}
\end{figure}

The physical interpretation can empower the learned model with good predictive ability.
Figure~\ref{fig:generalizability} presents the predicted lift force at different angles of attack from~$1^\circ$ to $18^\circ$.
The results show that the learned model can be well generalized to other flow conditions at different angles of attack.
It can be seen that the baseline model has significant discrepancies in the lift coefficients for angles of attacks larger than around $7.5^\circ$.
In contrast, the learned model improves the prediction on $C_l$ across the angle~$\alpha$.
The reason for the improved prediction in other angles is due to the appropriate estimation of flow separation, as shown at the top of Figure~\ref{fig:generalizability}.
Specifically, at the small angle of attack, e.g., $\alpha=1^\circ$, the baseline and learned models produce similar attached flow around the airfoil, and hence both predict the lift force in good agreement with the experiment.
However, at the large angle of attack $\alpha=11^\circ$, the baseline model still expects attached flow, which leads to the overestimation in $C_l$ compared to the experimental data.
In contrast, the learned model captures the flow separation near the trailing edge, which reduces the lift force and provides a good agreement with the lift force data.
Additionally, at $\alpha=18^\circ$, the baseline model predicts the flow separation but still underestimates the separation bubble size, which leads to the lift force being larger than the experimental measurement.
The learned model leads to the early onset of the flow separation and provides a larger separation bubble compared to the baseline model, thereby improving the lift force prediction.
The predictive performance of the learned model at additional angles $\alpha=11^\circ$ and $18^\circ$ can be found in~\ref{sec:unseen}.
Further generalization tests with different geometries are beyond the scope of the present work and will be conducted in the near future.
We note that the adjoint-based learning method has been used in the S809 airfoil, demonstrating that the learned model can be generalized well for different configurations such as the S805 and S814 airfoils~\cite{singh2017machine}.
Here we use the ensemble Kalman method which is comparable to the adjoint method in model learning as demonstrated in Ref.~\cite{zhang_ensemble-based_2022}.
Hence the current learned model could be generalized to other cases as the adjoint method does.

The S809 airfoil case has been used in various works~\cite{singh2017machine,holland2019field,Heo2022,wang2023unified}, including Singh et al. (2017)~\cite{singh2017machine}, where the learned model suppresses the turbulent production at the upstream boundary layer, leading to early flow separation.
The difference between the present work and previous studies mainly lies in two aspects.
First, the current modeling framework and the training method are different from the previous works. 
Specifically, the nonlinear eddy viscosity model is used in this study since it is flexible to capture both separated and secondary flows, while previous studies including Singh et al. (2017)~\cite{singh2017machine} use a linear eddy viscosity model which is not able to predict secondary flows.
Moreover, the ensemble Kalman method is used to train the neural networks in this work, while  previous studies~\cite{singh2017machine,holland2019field,Heo2022} often use the adjoint-based method for model inference.
Second, the case in this work highlights the consistency of the learned model behaviors independent of the training techniques, in addition to the discussion on the capability of the data-driven method in prediction improvement as was only done in previous studies.
That is, different model representations and training methods lead to similar predictive improvements and model behaviors.

\subsection{Flow in a square duct}

\subsubsection{Training performance}
\begin{figure}[!htb]
    \centering
    \includegraphics[width=0.65\textwidth]{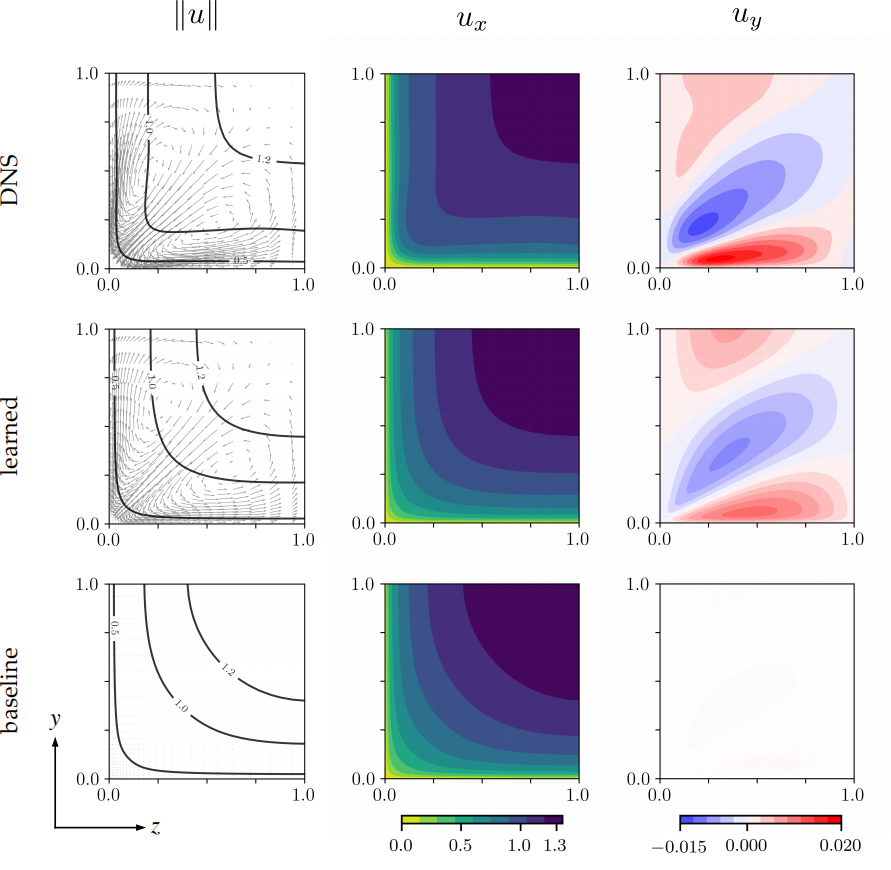}
    \caption{
    Velocity~$u_x$ and $u_y$ predicted from the learned models (center row) and baseline model (bottom row), compared with the ground truth (top row), for the square duct case. The velocity vectors are plotted along with contours of the streamwise velocity $u_x$.
    }
    \label{fig:sd_results_u}
\end{figure}

The ensemble Kalman method can learn a neural network-based turbulence model with improved velocity prediction for the square duct case.
It can be seen from Figure~\ref{fig:sd_results_u} where the vector and contour plots of velocities are presented.
The vector plots are presented in the first column of Figure~\ref{fig:sd_results_u}, where the isolines indicate the levels of $u_y=0.5, 1.0$ and $1.2$. 
It shows that the baseline model cannot predict the in-plane secondary flow, while the ensemble-based learned method can estimate the in-plane velocity vectors in a similar pattern as the DNS data.
The contour plots of $u_x$ and $u_y$ are presented in the last two columns of Figure~\ref{fig:sd_results_u}.
The plots of $u_z$ are omitted for brevity since it is symmetric to the vertical velocity~$u_y$.
The axial velocity~$u_x$ with the baseline model and the learned model both have good agreement with the DNS data.
The vertical velocity~$u_y$ is not captured at all by the baseline model, while the learned model can capture similar patterns as the DNS results.

\begin{figure}[!htb]
    \centering
    \begin{tabular}{ccccc}
        &  $\tau_{yy}$ & $\tau_{zz}$ & $\tau_{yy}-\tau_{zz}$ & $\tau_{yz}$\\
        \rotatebox[origin=c]{90}{DNS} & 
        \raisebox{-.5\height}{\includegraphics[scale=0.5]{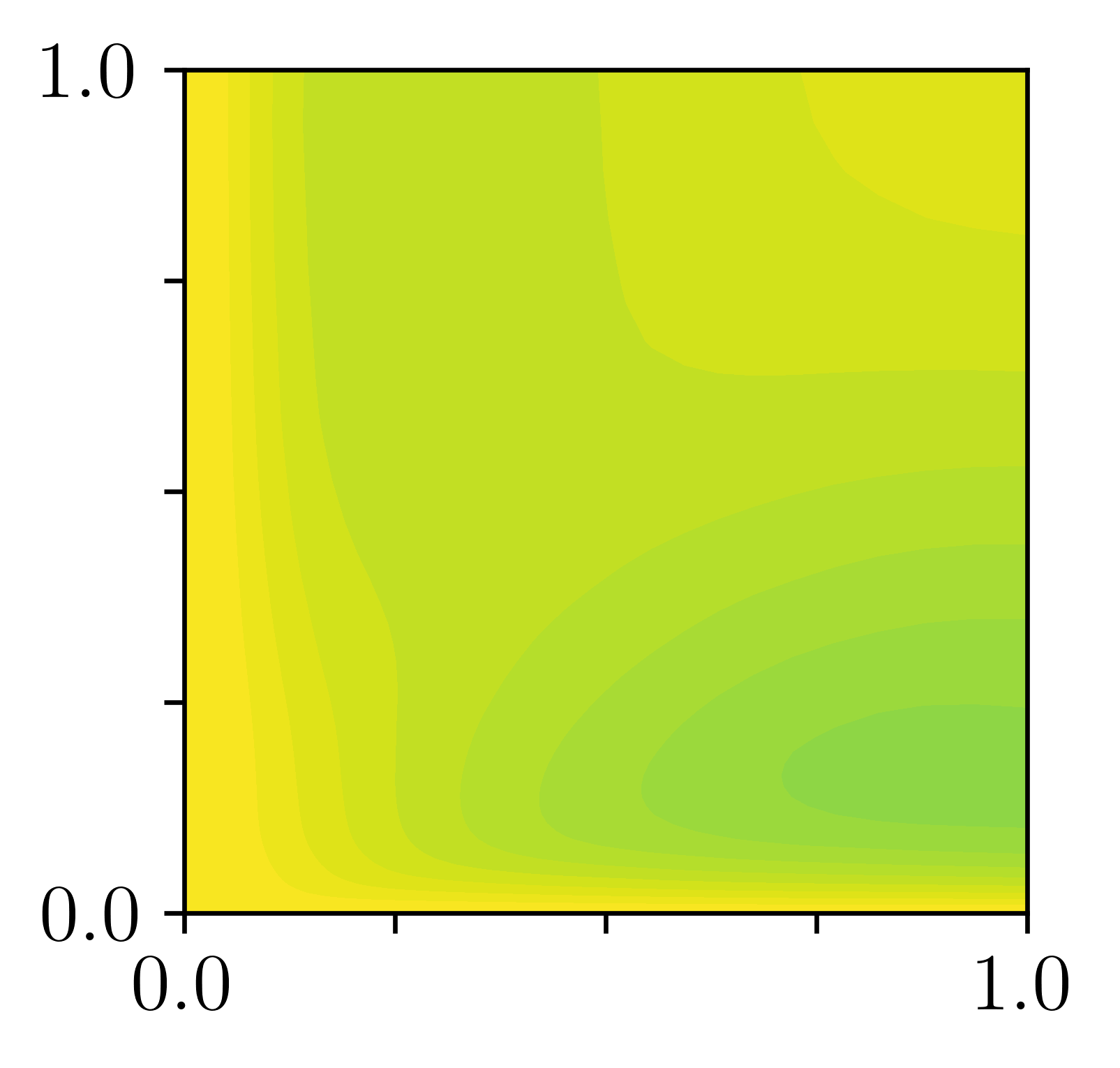}} &
        \raisebox{-.5\height}{\includegraphics[scale=0.5]{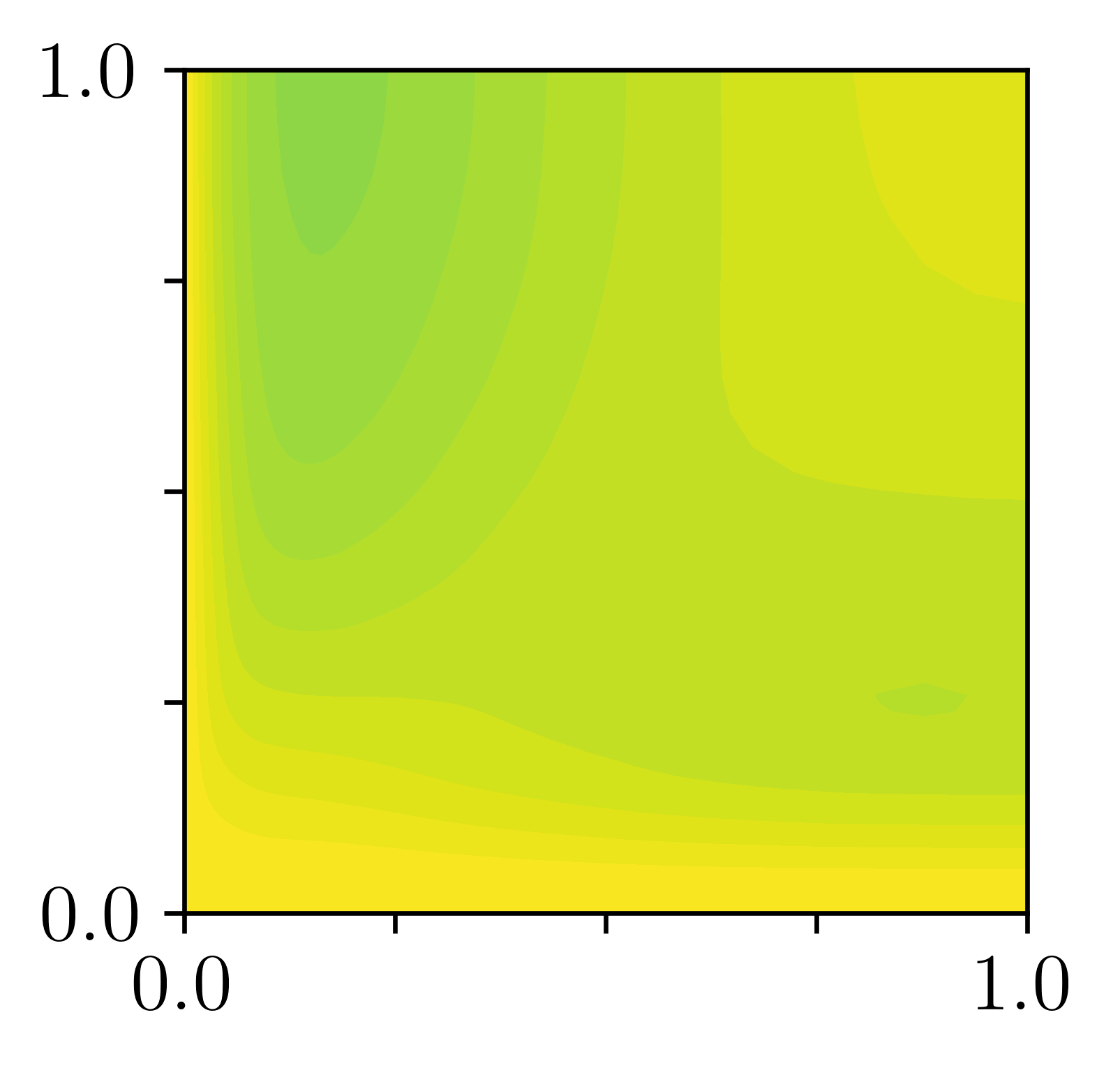}} &
        \raisebox{-.5\height}{\includegraphics[scale=0.5]{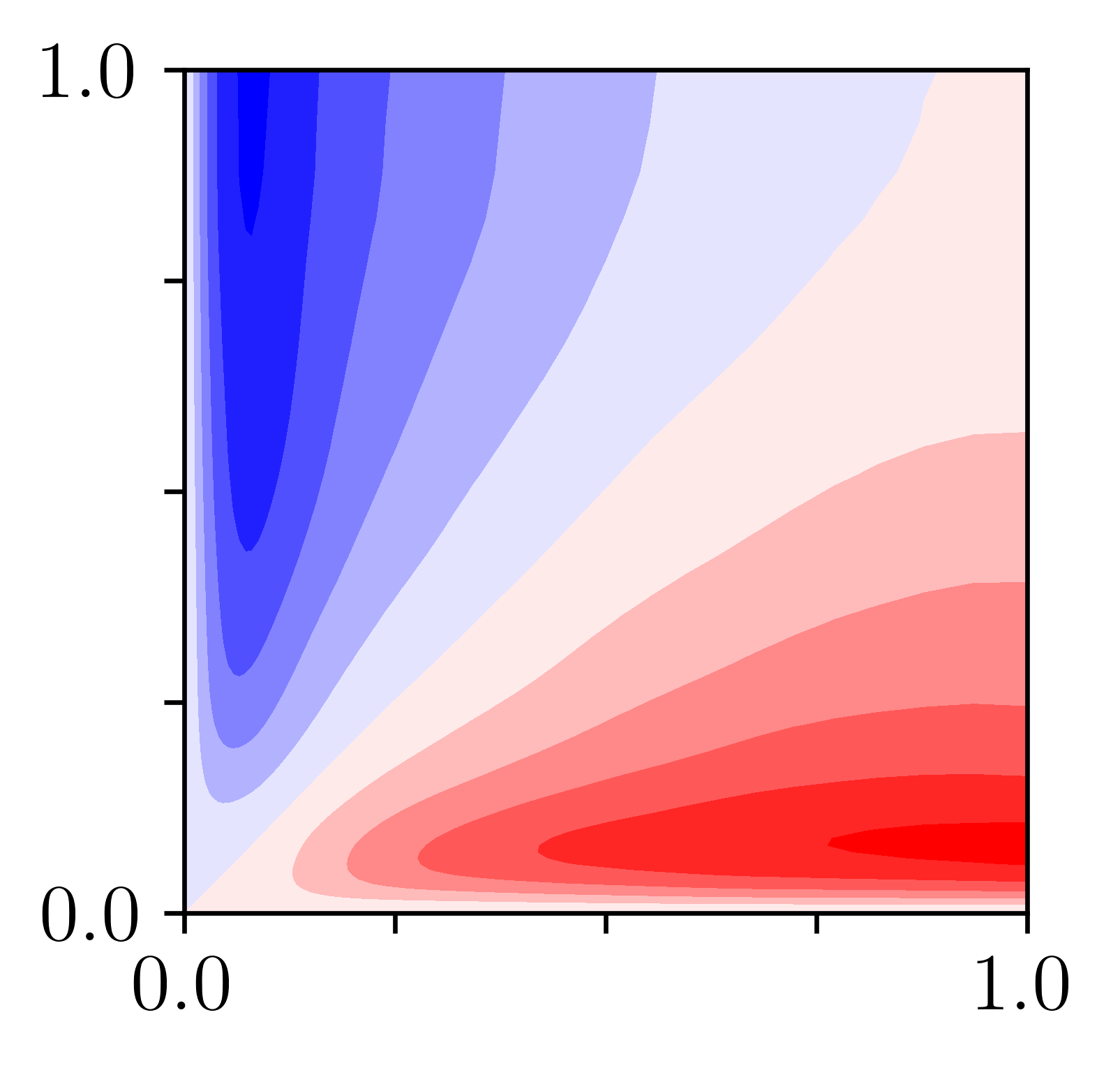}} &
        \raisebox{-.5\height}{\includegraphics[scale=0.5]{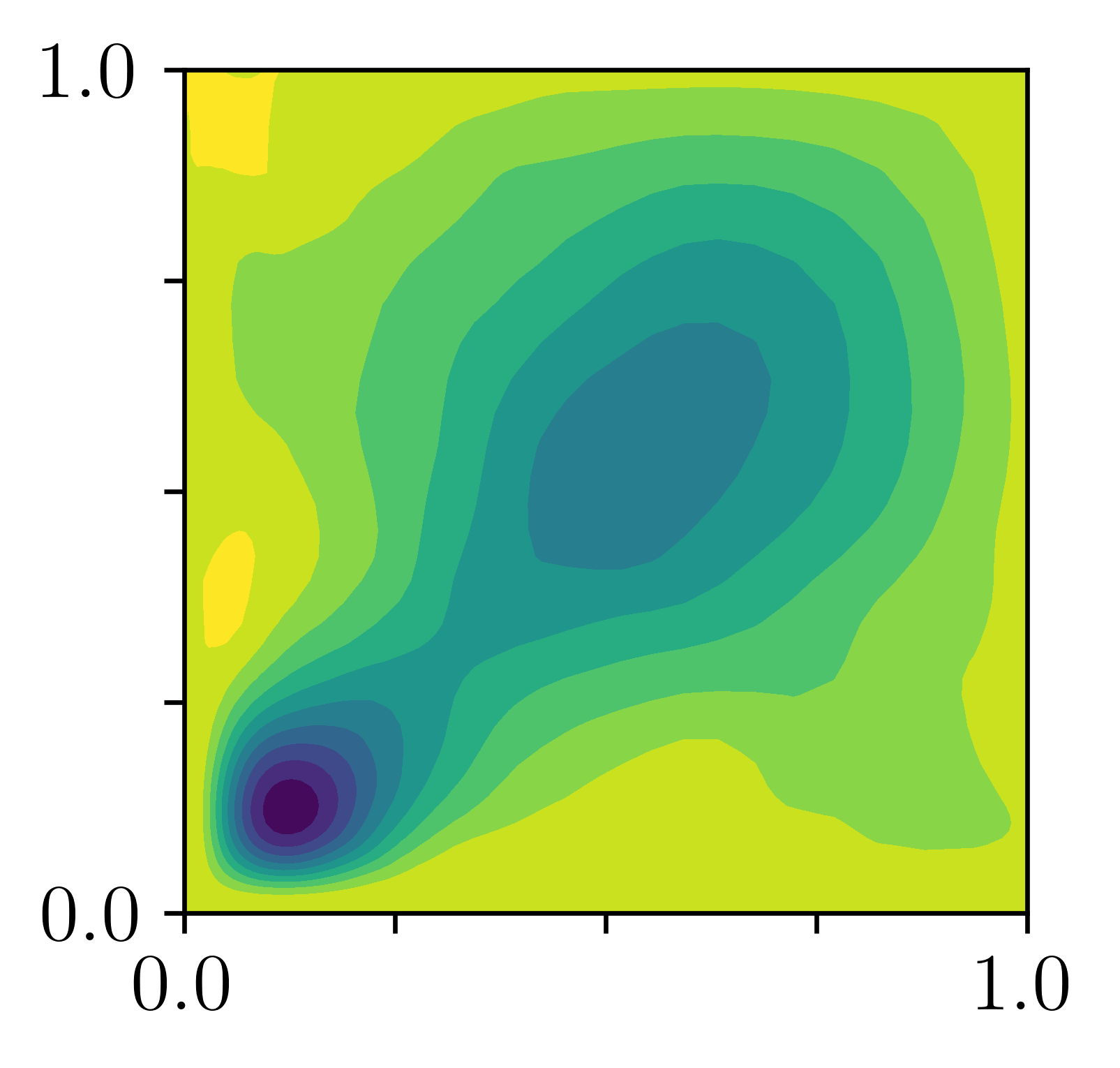}} 
        \\
        \rotatebox[origin=c]{90}{learned} & 
        \raisebox{-.5\height}{\includegraphics[scale=0.5]{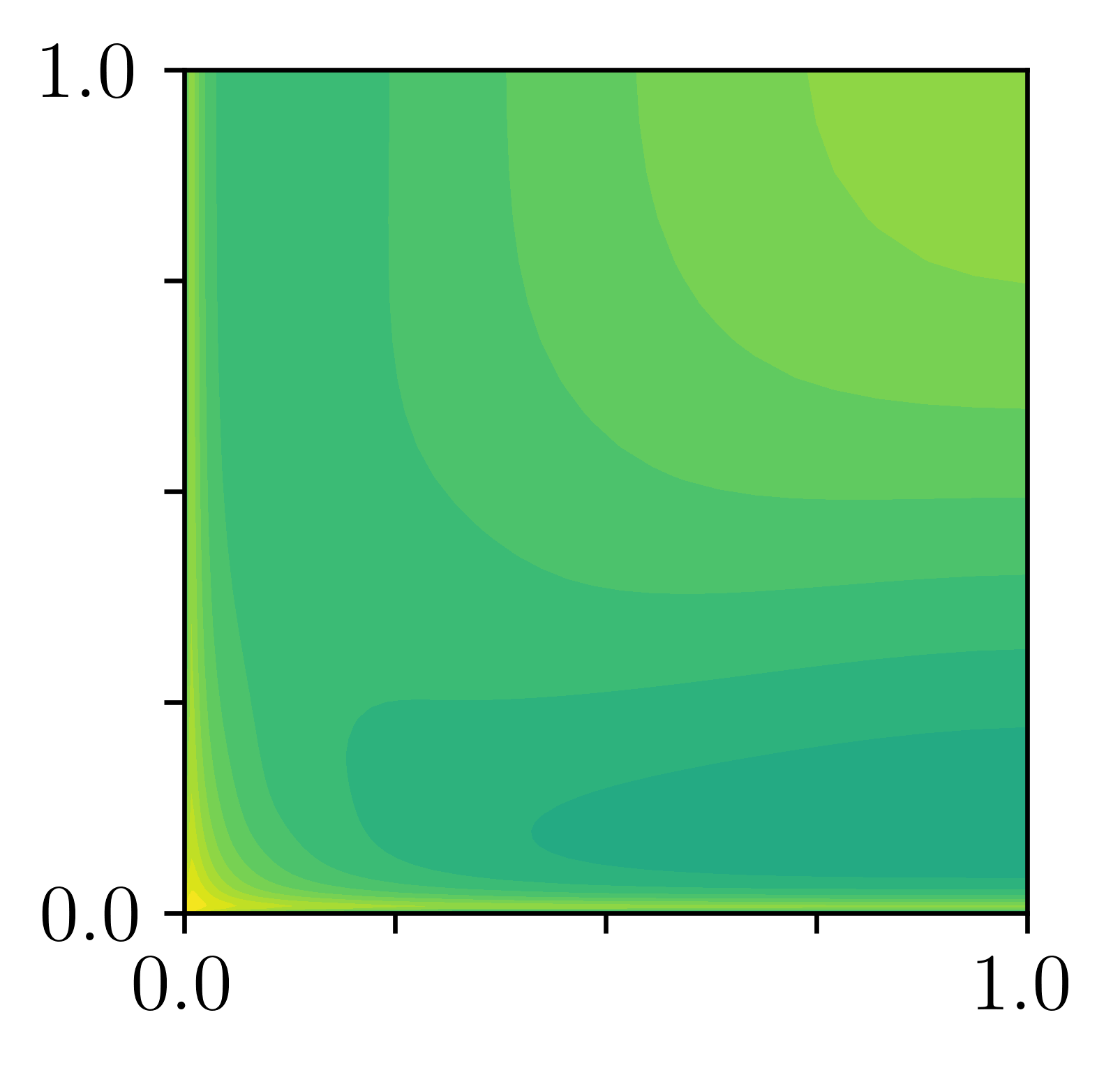}} &
        \raisebox{-.5\height}{\includegraphics[scale=0.5]{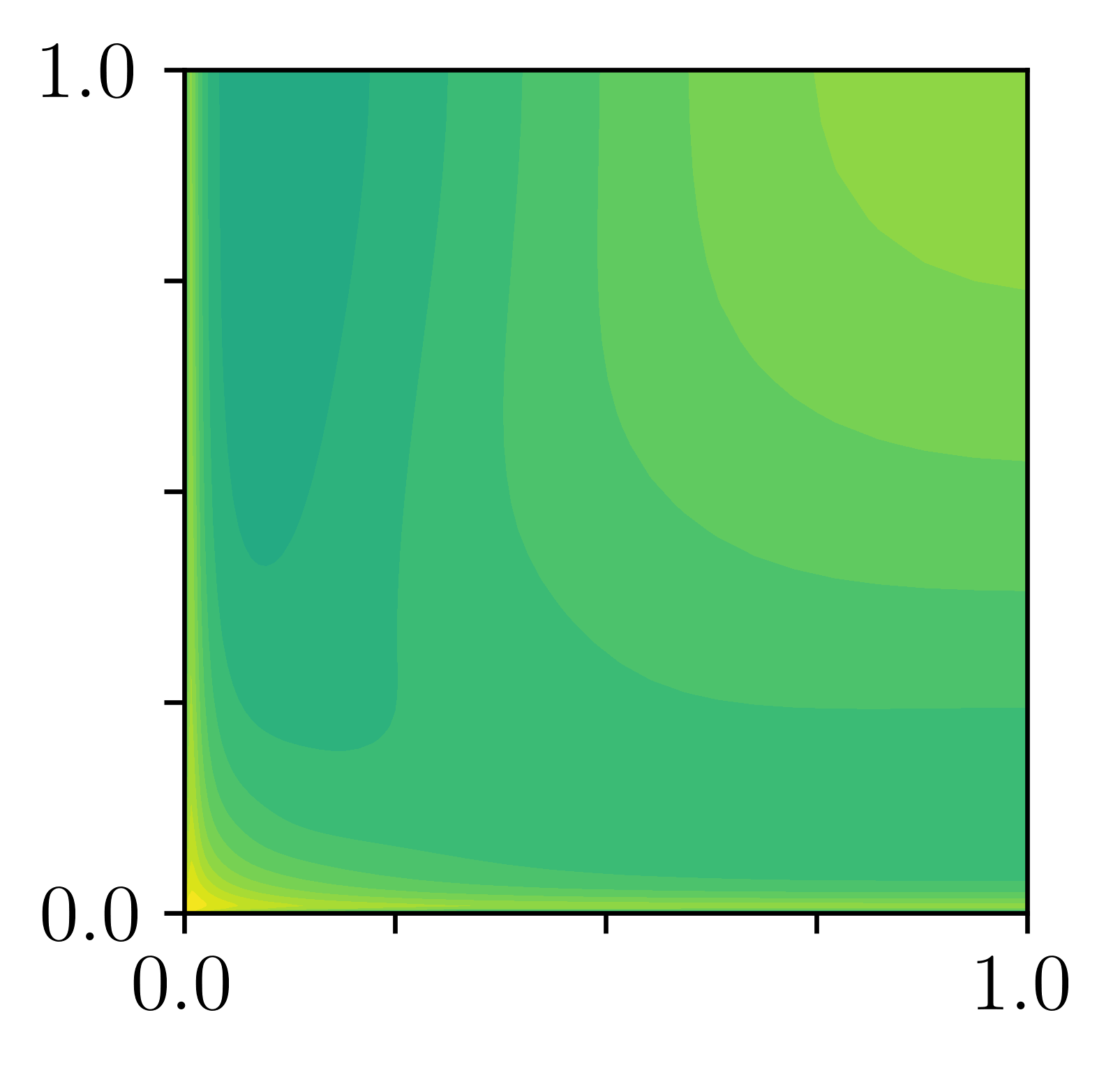}} &
        \raisebox{-.5\height}{\includegraphics[scale=0.5]{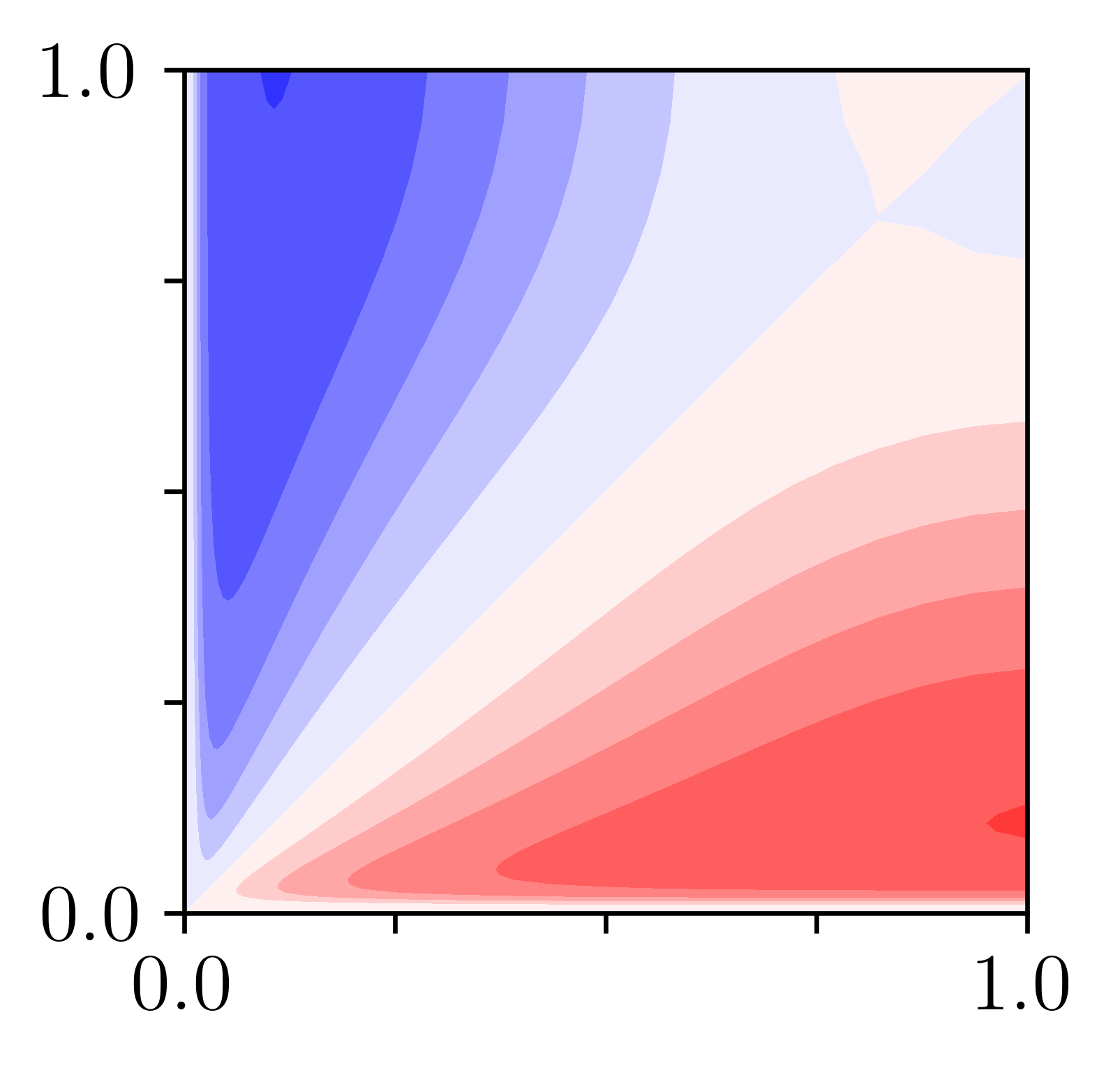}} &
        \raisebox{-.5\height}{\includegraphics[scale=0.5]{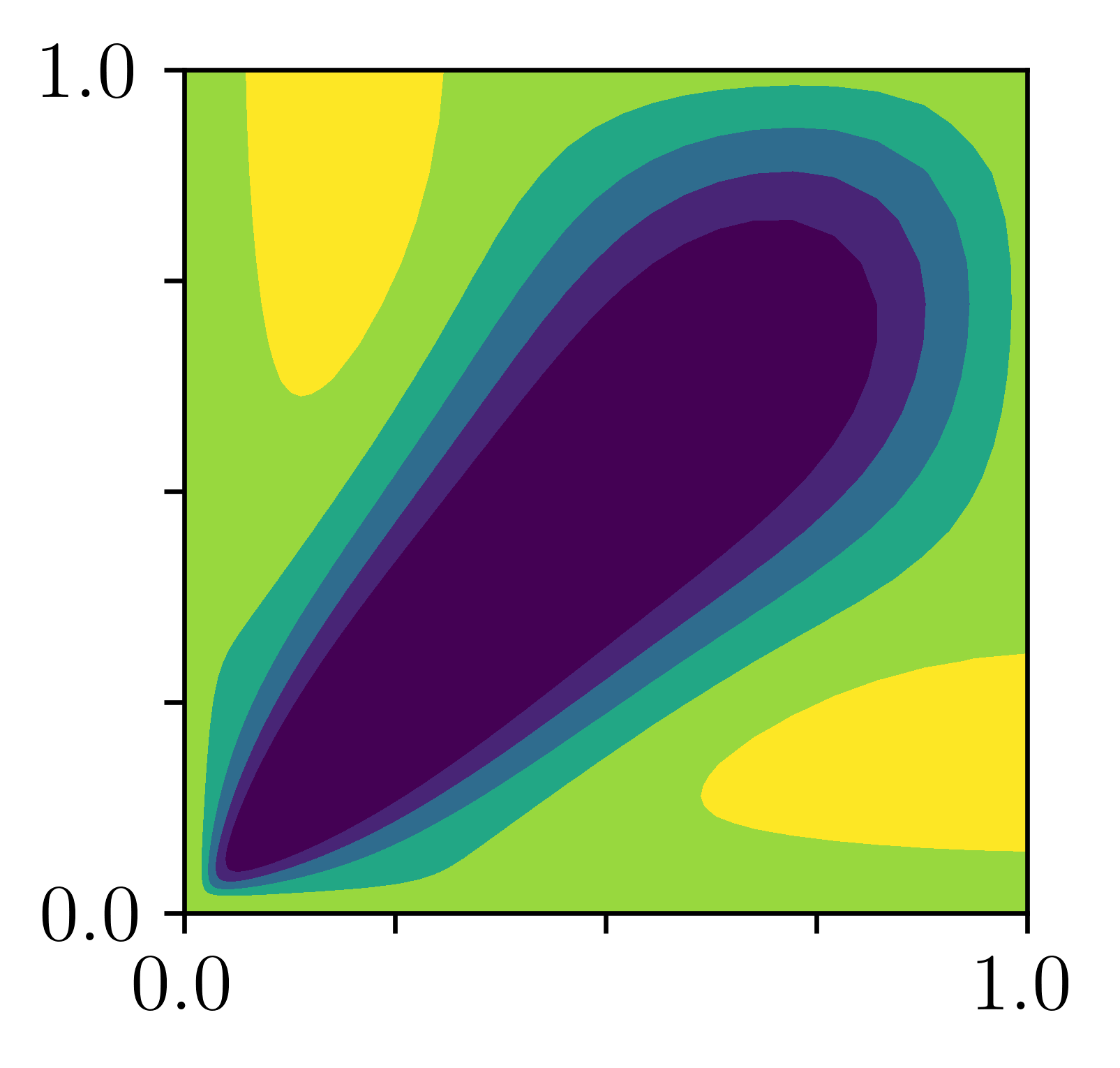}} 
        \\
        \rotatebox[origin=c]{90}{baseline} &
        \raisebox{-.5\height}{\includegraphics[scale=0.5]{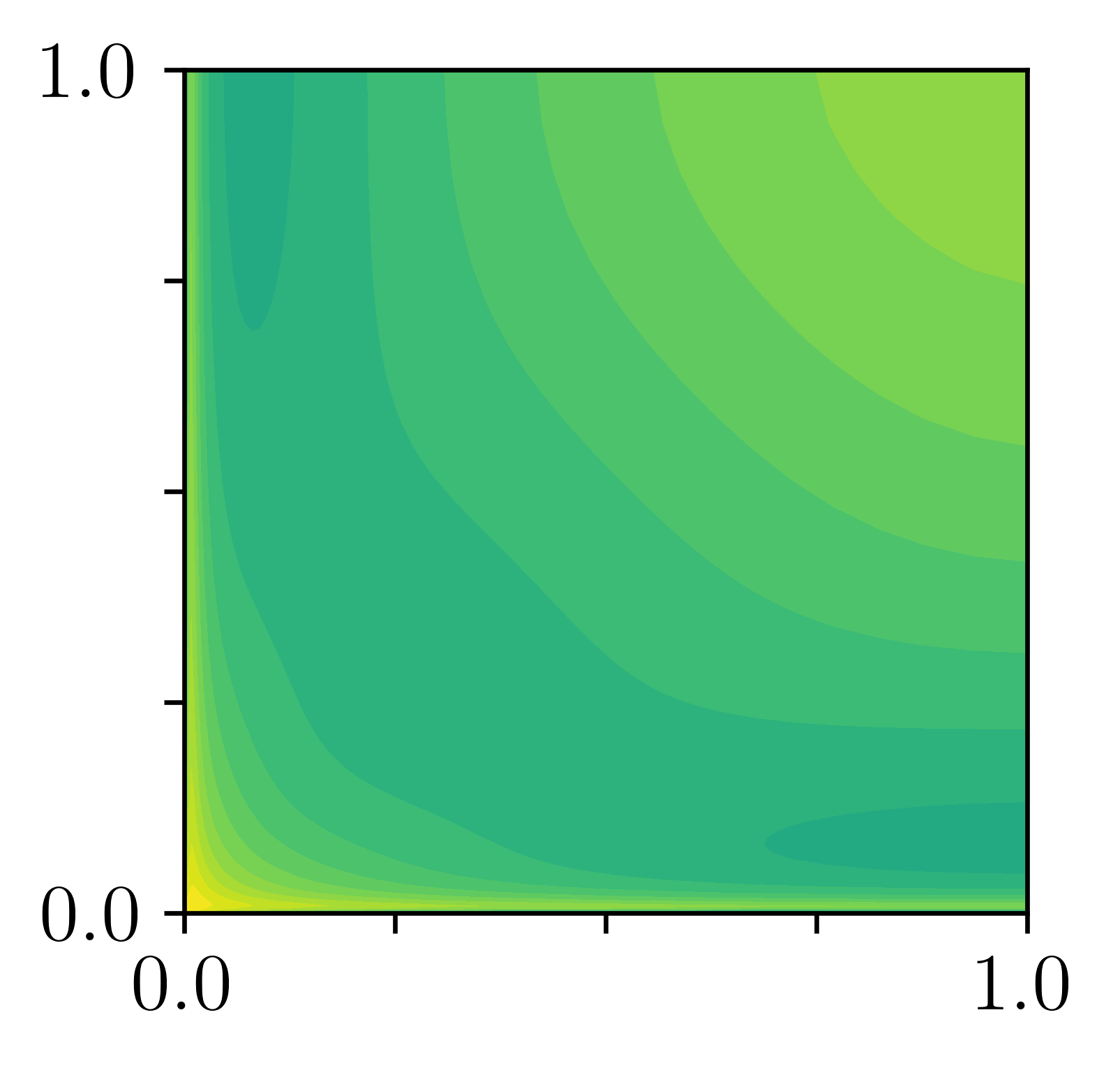}} &
        \raisebox{-.5\height}{\includegraphics[scale=0.5]{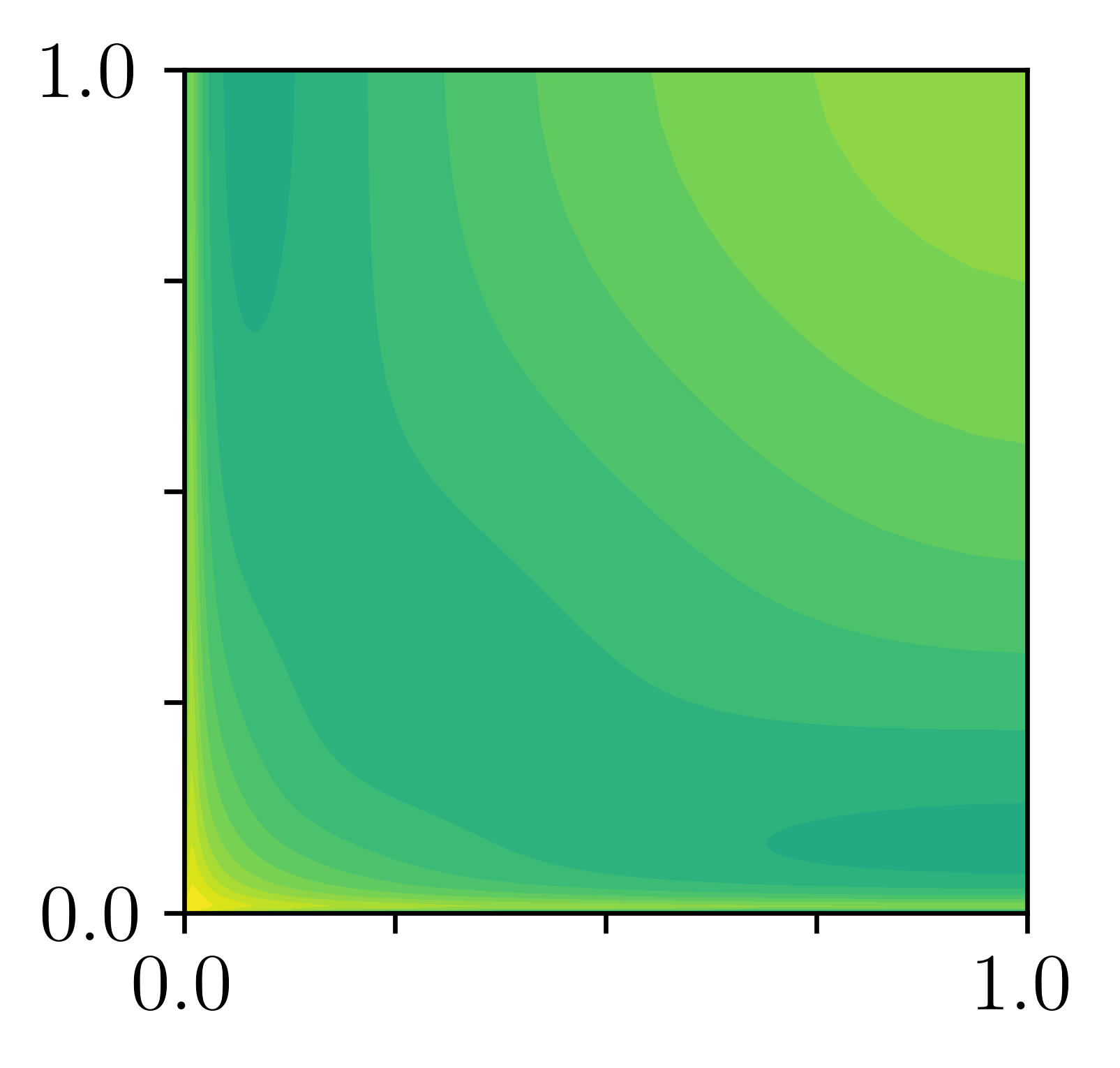}} &
        \raisebox{-.5\height}{\includegraphics[scale=0.5]{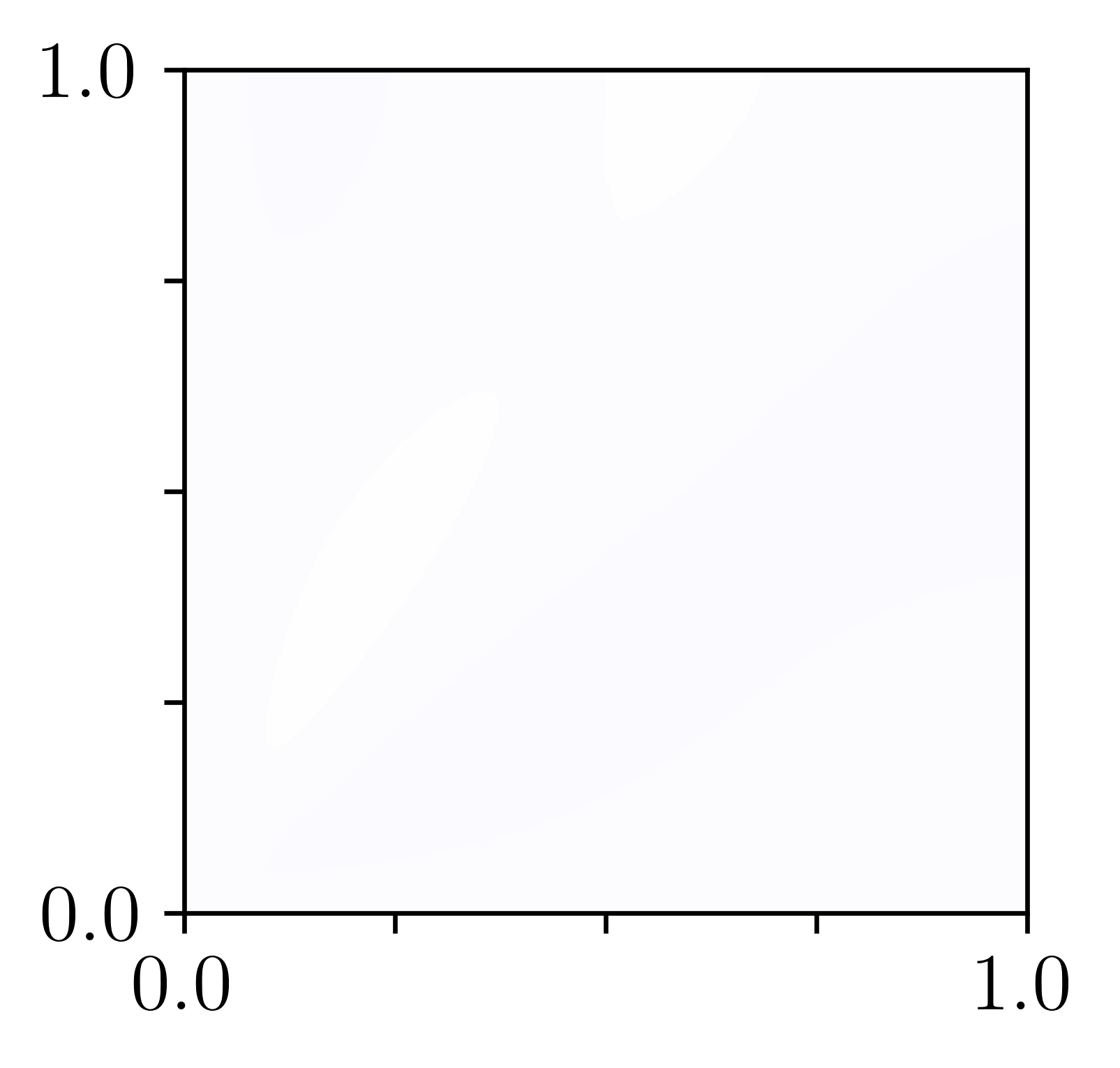}} &
        \raisebox{-.5\height}{\includegraphics[scale=0.5]{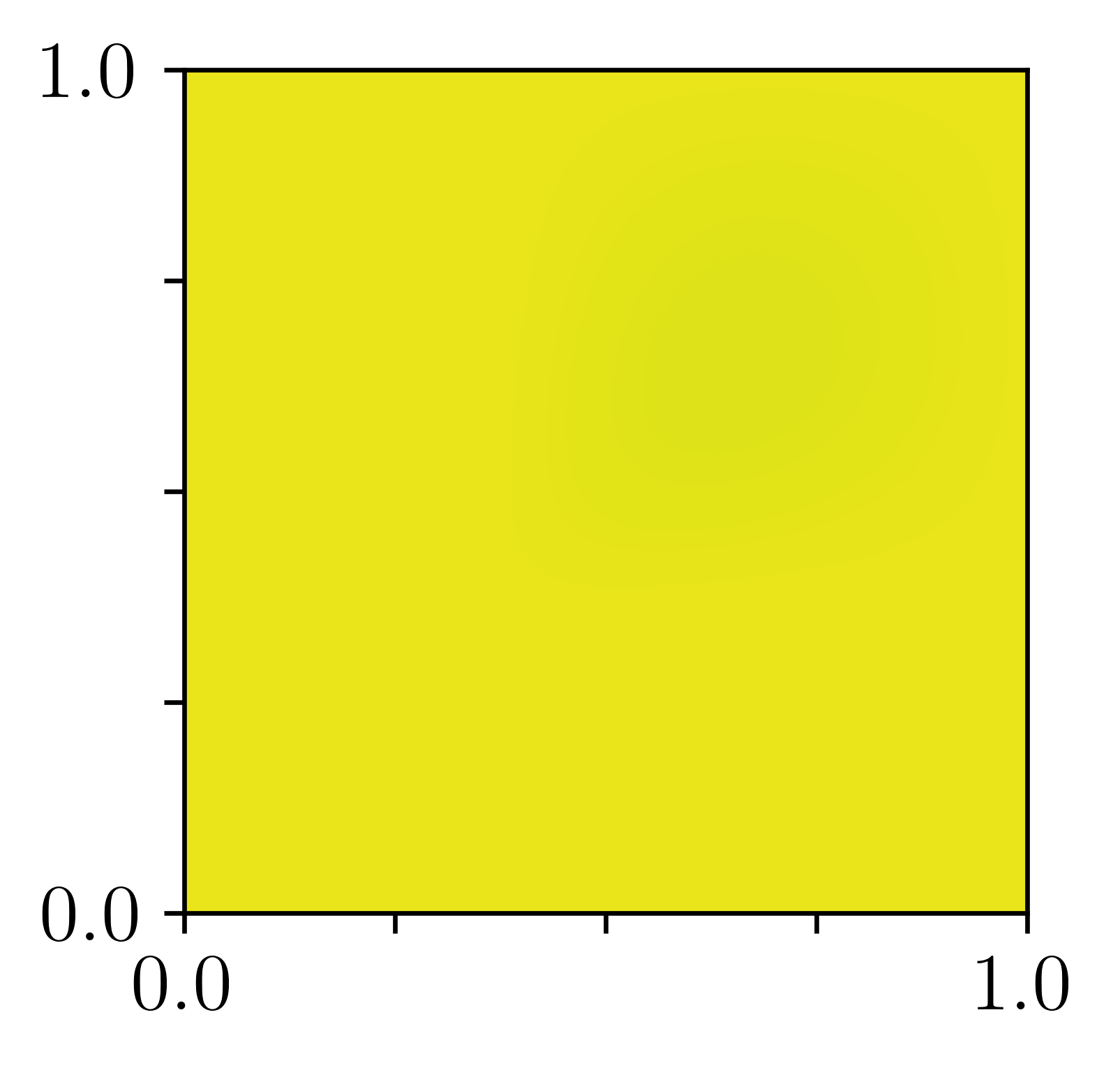}} 
        \\
        & \raisebox{-.5\height}{\includegraphics[scale=0.7]{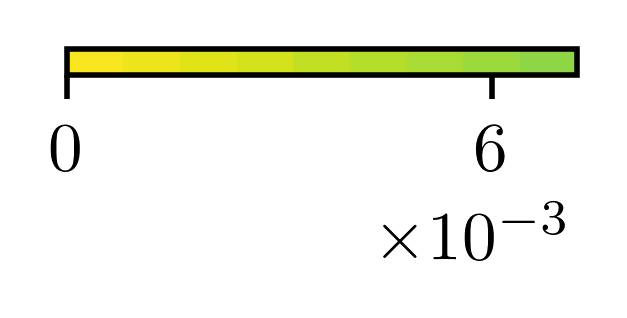}} 
        & \raisebox{-.5\height}{\includegraphics[scale=0.7]{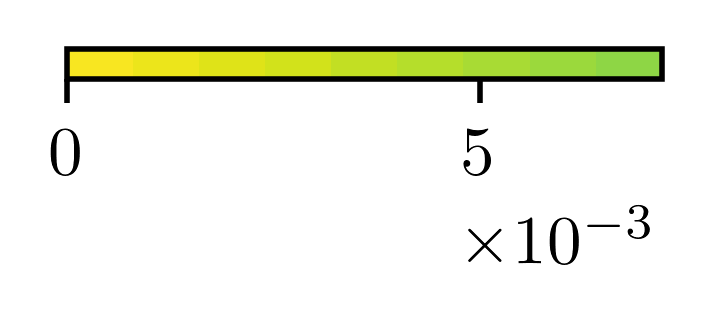}} 
        &\raisebox{-.5\height}{\includegraphics[scale=0.7]{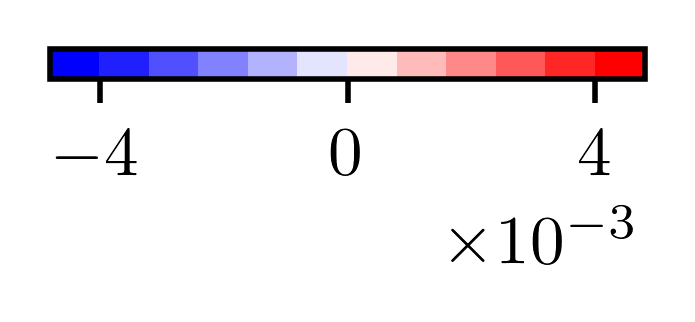}}
        & \raisebox{-.5\height}{\includegraphics[scale=0.7]{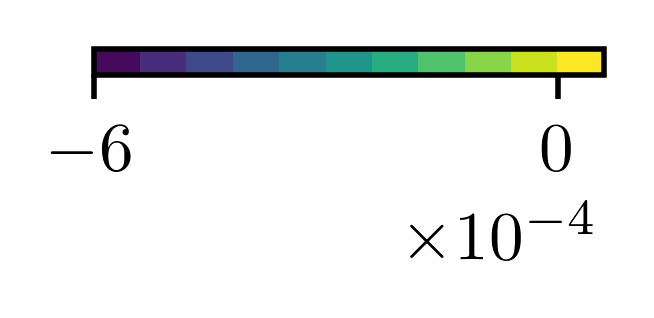}}
    \end{tabular}
    \caption{ 
     Reynolds normal stresses $\tau_{yy}$ and $\tau_{zz}$, Reynolds shear stresses $\tau_{yz}$, and imbalance of Reynolds normal stresses $\tau_{yy}-\tau_{zz}$ predicted from the learned model (center row) and the baseline model (bottom row), compared with the ground truth DNS (top row), for the square duct case. 
     }
    \label{fig:sd_results_tau}
\end{figure}

The learned model improves the prediction of the in-plane velocity by capturing the Reynolds stress imbalance and Reynolds shear stress.
It is supported by Figure~\ref{fig:sd_results_tau} where the Reynolds stress components and the imbalance of Reynolds normal stress are presented.
The in-plane velocity is driven by the Reynolds stress imbalance~$\tau_{yy} - \tau_{zz}$ and the Reynolds shear stress~$\tau_{yz}$ based on the axial vorticity transport equation~\cite{speziale_turbulent_1982}: 

\begin{equation}
    u_y \frac{\partial \omega_x}{\partial y}+u_z \frac{\partial \omega_x}{\partial z}-\nu \nabla^2 \omega_x+\frac{\partial^2}{\partial y \partial z}\left(\boldsymbol{\tau}_{z z}-\boldsymbol{\tau}_{y y}\right)+\left(\frac{\partial^2}{\partial y \partial y}-\frac{\partial^2}{\partial z \partial z}\right) \boldsymbol{\tau}_{y z} = 0 \text{.}
\end{equation}
For this reason, capturing the in-plane velocity requires well estimating  Reynolds normal stress imbalance~$\tau_{yy} - \tau_{zz}$ and Reynolds shear stress~$\tau_{yz}$.
From Figure~\ref{fig:sd_results_tau}, the learned model shows significant improvements in the prediction of $\tau_{yz}$ and $\tau_{yy}-\tau_{zz}$ compared to the baseline.
Although the model still has discrepancies with the DNS data near the duct center, the noticeable improvement in $\tau_{yz}$ and $\tau_{yy}-\tau_{zz}$ allow us to obtain good agreement to the DNS data in the in-plane velocity prediction.
Specifically, the baseline model estimates almost zero for both the Reynolds shear stress $\tau_{yz}$ and the imbalance of Reynolds normal stresses~$\tau_{yy} - \tau_{zz}$ in the entire computational domain.
In contrast, the learned model can predict them similarly obtained in the DNS, which significantly improves the in-plane velocity as shown in Fig.~\ref{fig:sd_results_u}.
For the Reynolds normal stresses~$\tau_{xx}$ and $\tau_{yy}$ and the Reynolds shear stress~$\tau_{xy}$, the baseline model and the learned model give similar predictions since the in-plane velocity can not guide the training in these Reynolds stress components.

\begin{figure}[!htb]
    \centering
    \includegraphics[width=0.6\textwidth]{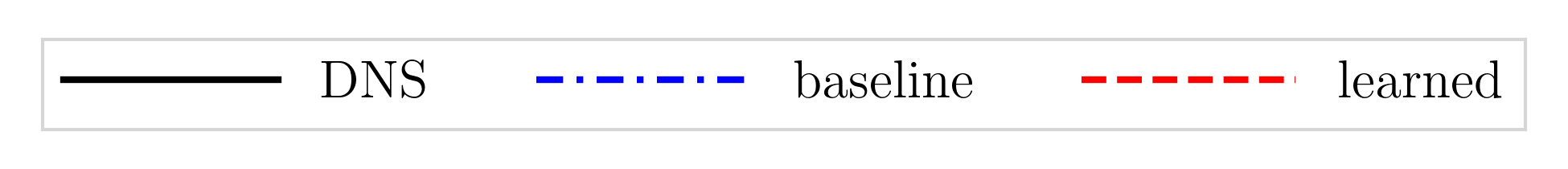} \\
    \subfloat[Velocity $u_{x}$ profiles]{\includegraphics[height=0.36\textwidth]{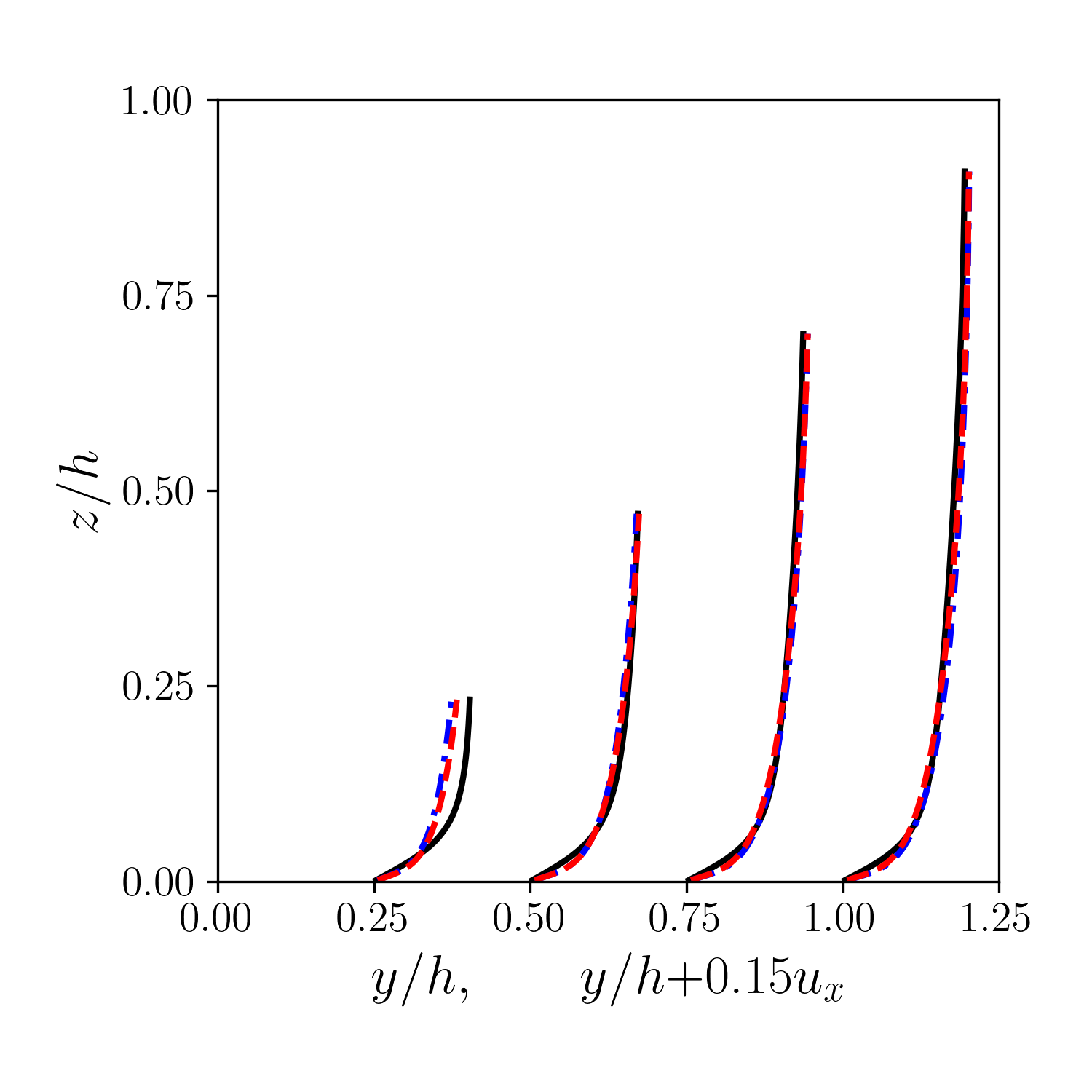}}
    \quad
    \subfloat[Velocity $u_{y}$ profiles]{\includegraphics[height=0.36\textwidth]{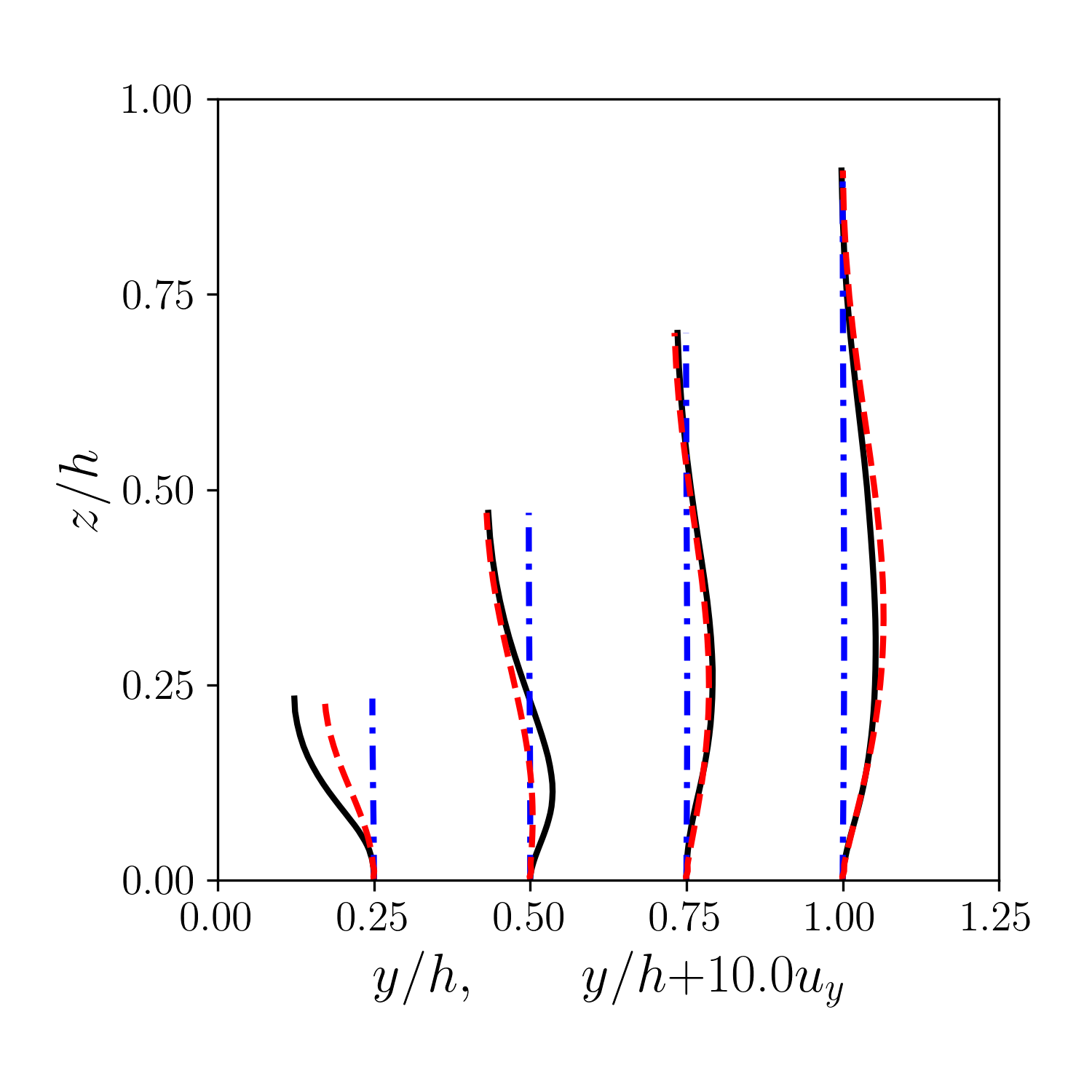}}
    \quad
    \subfloat[Reynolds Stress $\tau_{yy}-\tau_{zz}$ profiles]{\includegraphics[height=0.36\textwidth]{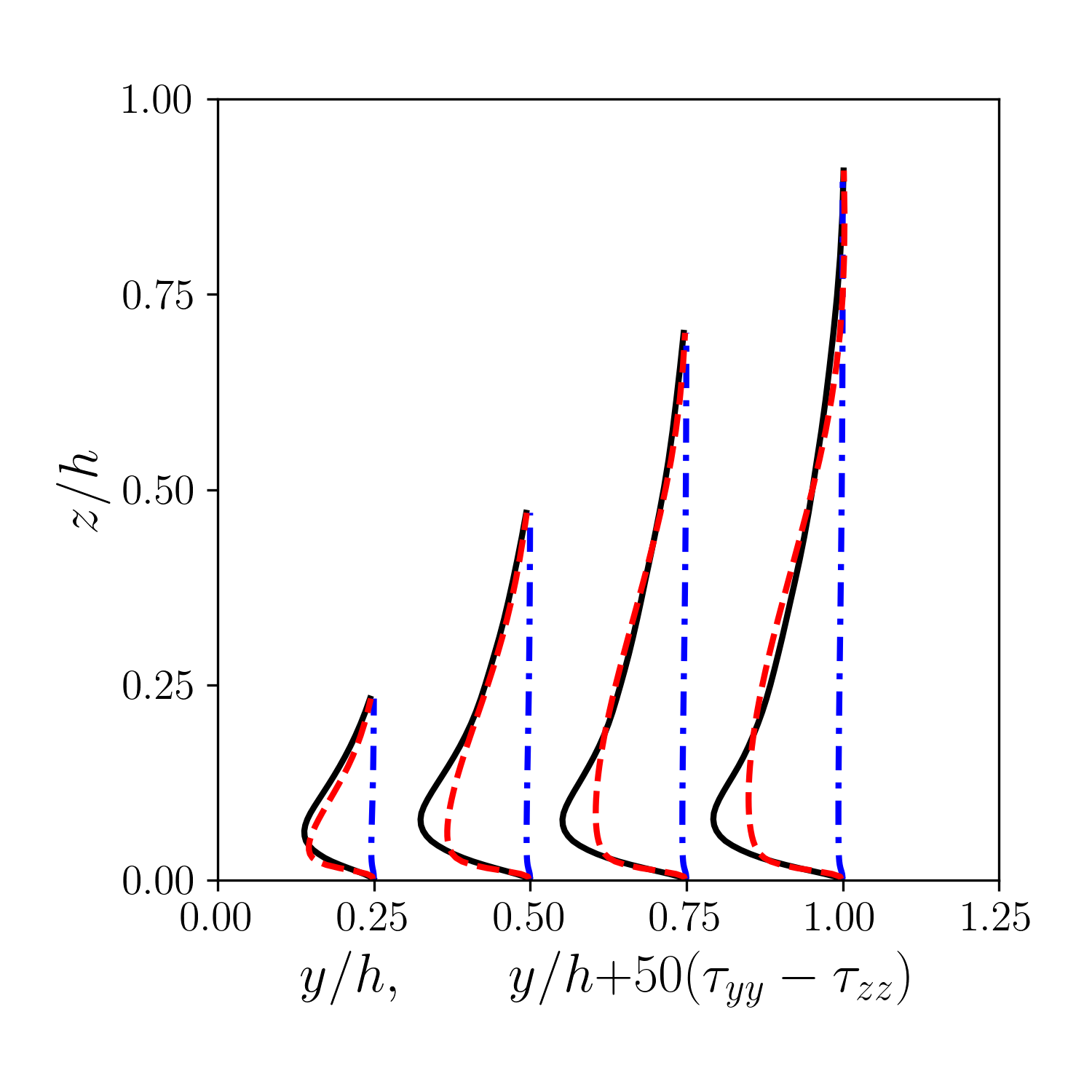}}
    \quad
    \subfloat[Reynolds Stress $\tau_{yz}$ profiles]{\includegraphics[height=0.36\textwidth]{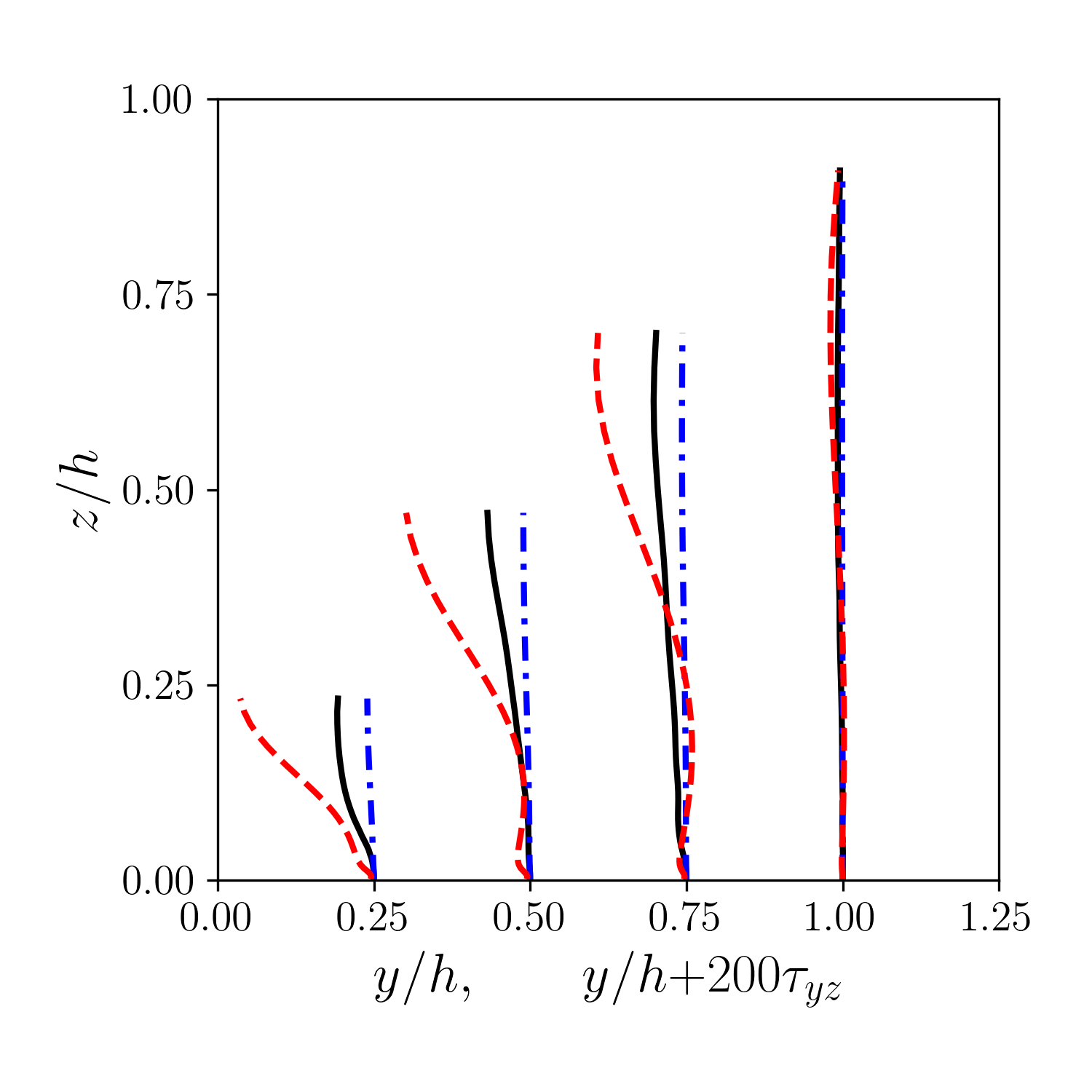}}
    \caption{Prediction of velocity and Reynolds stress along profiles at $y/H=0.25, 0.5, 0.75, 1$ with comparison among the learned model, the baseline model, and the experimental data, for the square duct case
    }
    \label{fig:sd_results_tau_profiles}
\end{figure}

The profiles of velocity and the Reynolds stress at $y/h=0.25, 0.5, 0.75, 1$ are provided in Fig.~\ref{fig:sd_results_tau_profiles}.
It can be seen that the streamwise velocity~$u_x$ is similar between the baseline model and the learned model, and both can have good agreement with the DNS data.
As for the in--plane velocity~$u_y$, the baseline model is not able to predict the in-plane velocity and provide $u_y=0$ at the entire domain.
In contrast, the learned model significantly improves the prediction of $u_y$ in better agreement with the DNS data.
The plots of the Reynolds stress show that the learned model provides better predictions in the imbalance of the Reynolds normal stress~$\tau_{yy} - \tau_{zz}$ than the baseline model.
As for the Reynolds shear stress~$\tau_{yz}$, both the learned and baseline models have noticeable discrepancies from the DNS data. 
Also, it is observed that the learned model has larger discrepancies near the diagonal line of the computational domain compared to the baseline model, which is consistent with the plots in Fig.~\ref{fig:sd_results_tau}.
Additional results of the velocity and the Reynolds stresses at $y/h=0.2, 0.4, 0.6, 0.8$ are presented in~\ref{sec:unseen}.

\subsubsection{Physical interpretation of model behavior}
The behavior of the learned model can be interpreted based on the learned tensor coefficients~$g$.
In the secondary flow, the axial velocity $u_x$ is orders of magnitude larger than the in-plane velocity.
Also, only four Reynolds stress components, i.e., Reynolds shear stress~$\tau_{xy}$, $\tau_{xz}$, $\tau_{yz}$, and Reynolds normal stress imbalance~$\tau_{yy} - \tau_{zz}$, affect the velocity~\cite{michelen2021machine}.
The former two components affect the axial velocity, and the latter components of $\tau_{yz}$ and $\tau_{yy} - \tau_{zz}$ affect in-plane velocity.
It can be further derived~\cite{michelen2021machine} that only the coefficient~$g^{(1)}$ and the combination $g^{(2)} - 0.5 g^{(3)} + 0.5 g^{(4)}$ can be learned with velocity data in the scenario of only first four tensor bases.
Moreover, there is only one independent scalar invariant since $\theta_1 \approx - \theta_2$~\cite{strofer2021end}.
Therefore, we investigate the functional mapping from the scalar invariant~$\theta_1$ to the coefficient~$g^{(1)}$ and the combination~$g^{(2)} - 0.5 g^{(3)} + 0.5 g^{(4)}$.

\begin{figure}[!htb]
    \centering
    \subfloat[$g^{(1)}$ function]{\includegraphics[width=0.3\textwidth]{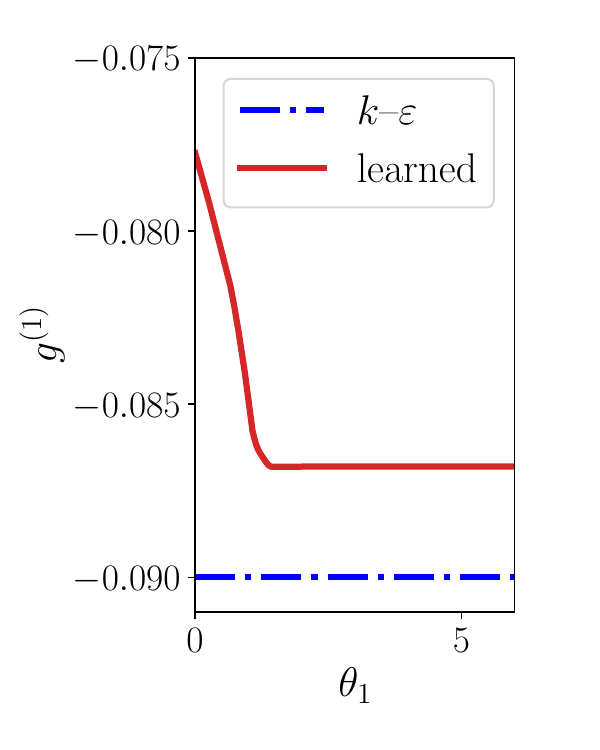}}
    \subfloat[$g^{(2)}-0.5g^{(3)}+0.5g^{(4)}$ function]{\includegraphics[width=0.3\textwidth]{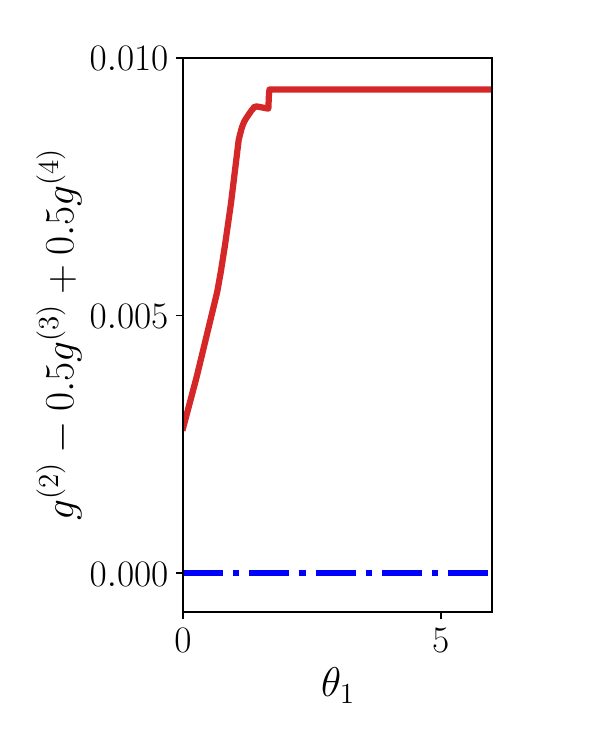}}
    \caption{Comparison of the model function $g^{(1)}$ and the combination~$g^{(2)}-0.5g^{(3)}+0.5g^{(4)}$ between the learned and the baseline models for the square duct case
    }
    \label{fig:gfuncs_duct}
\end{figure}

The coefficient $g^{(1)}$ and the combination of $g^{(2-4)}$ are shown in Figure~\ref{fig:gfuncs_duct}.
The learned function of $g^{(1)}$ can be seen from Figure~\ref{fig:gfuncs_duct}(a).
Note that the coefficient $g^{(1)}$ is equivalent to the $-C_\mu$ of the $k$--$\varepsilon$ model.
The difference lies in that the coefficient has dependencies on local scalar invariants in this work rather than a constant, i.e., $-0.09$.
The $g^{(1)}$ function with the learned model varies slightly from $-0.87$ to $-0.78$.
For the small scalar invariants that are located around the duct center, the magnitude of the $g^{(1)}$ function is less than $0.08$.
As the scalar invariant increases, the magnitude increases to around $0.087$, which is slightly less than the baseline value (i.e., 0.09).
The baseline model provides the combination of~$g^{(2-4)}$ at almost zero, which cannot capture the in-plane velocity.
In contrast, the learned model increases the magnitude of the combination at the range of nearly $[0.0025, 0.01]$.
This leads to nonlinear functional mappings between the Reynolds stress and the strain rate.
Such nonlinear models capture the Reynolds shear stress~$\tau_{yz}$ and the Reynolds normal stress imbalance~$\tau_{yy} - \tau_{zz}$, which further improve the prediction of the in-plane velocity. 

\begin{figure}[!htb]
    \centering
    \subfloat[$\| g^{(1)} \mathbf{T}^{(1)} \|$]{\includegraphics[width=0.2\textwidth]{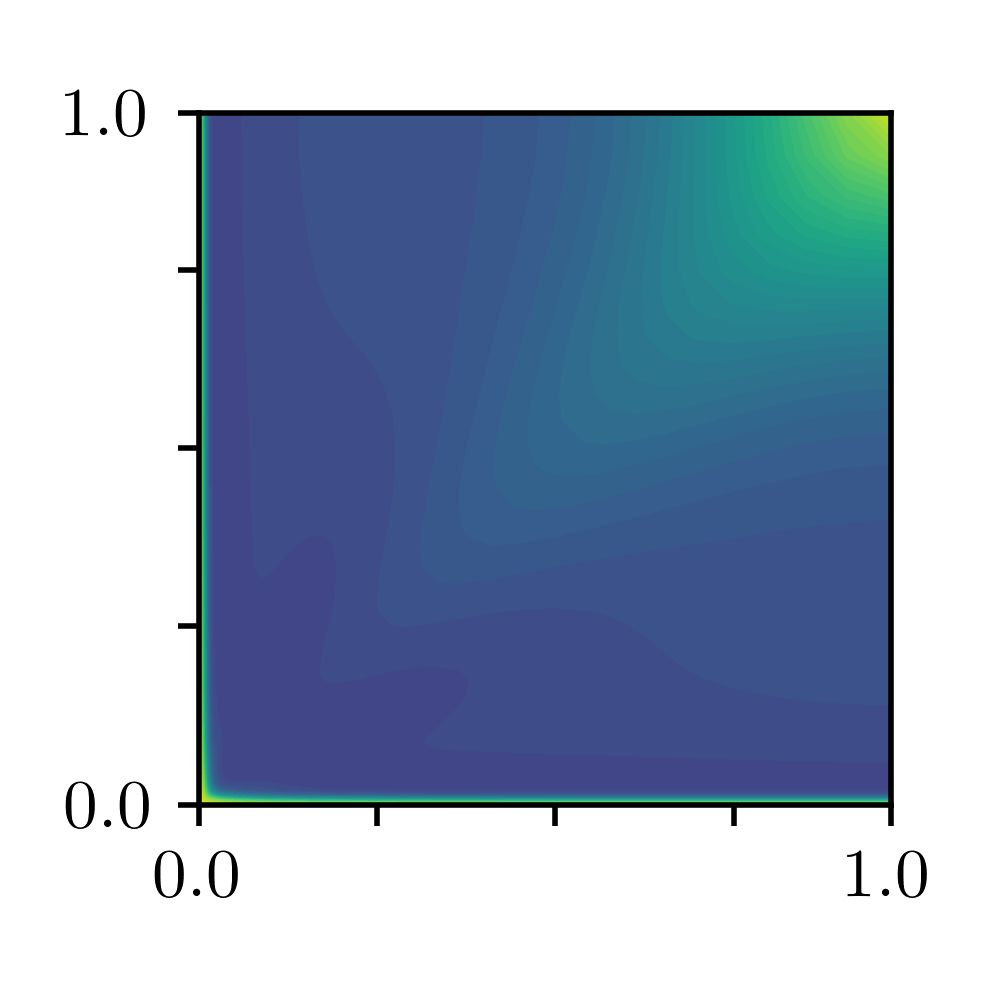}}
    \includegraphics[width=0.07
\textwidth,trim=0 -0.5cm 0 0]{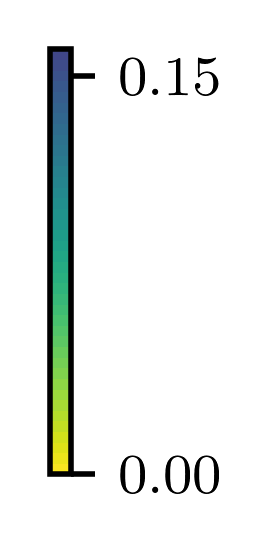}
    \subfloat[$\| g^{(2)} \mathbf{T}^{(2)} \|$]
    {\includegraphics[width=0.2\textwidth]{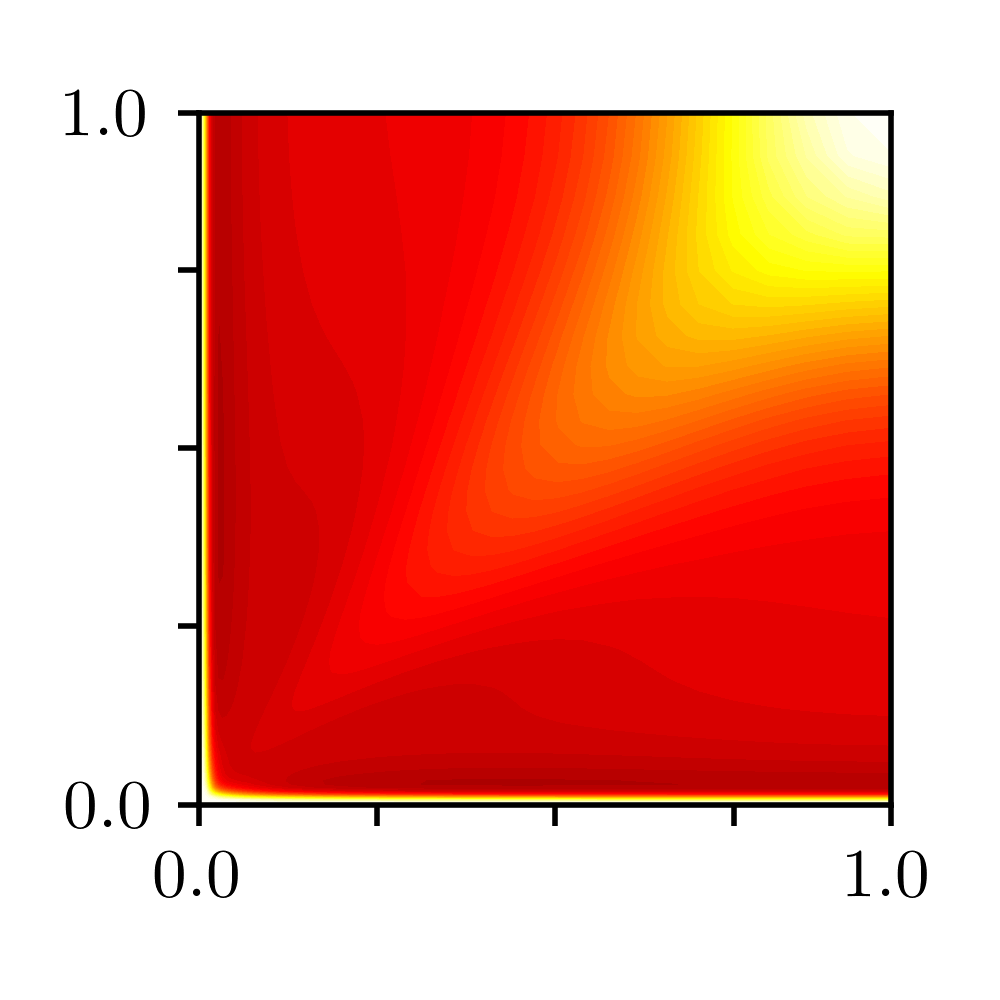}}
    \subfloat[$\| g^{(3)} \mathbf{T}^{(3)} \|$]
    {\includegraphics[width=0.2\textwidth]{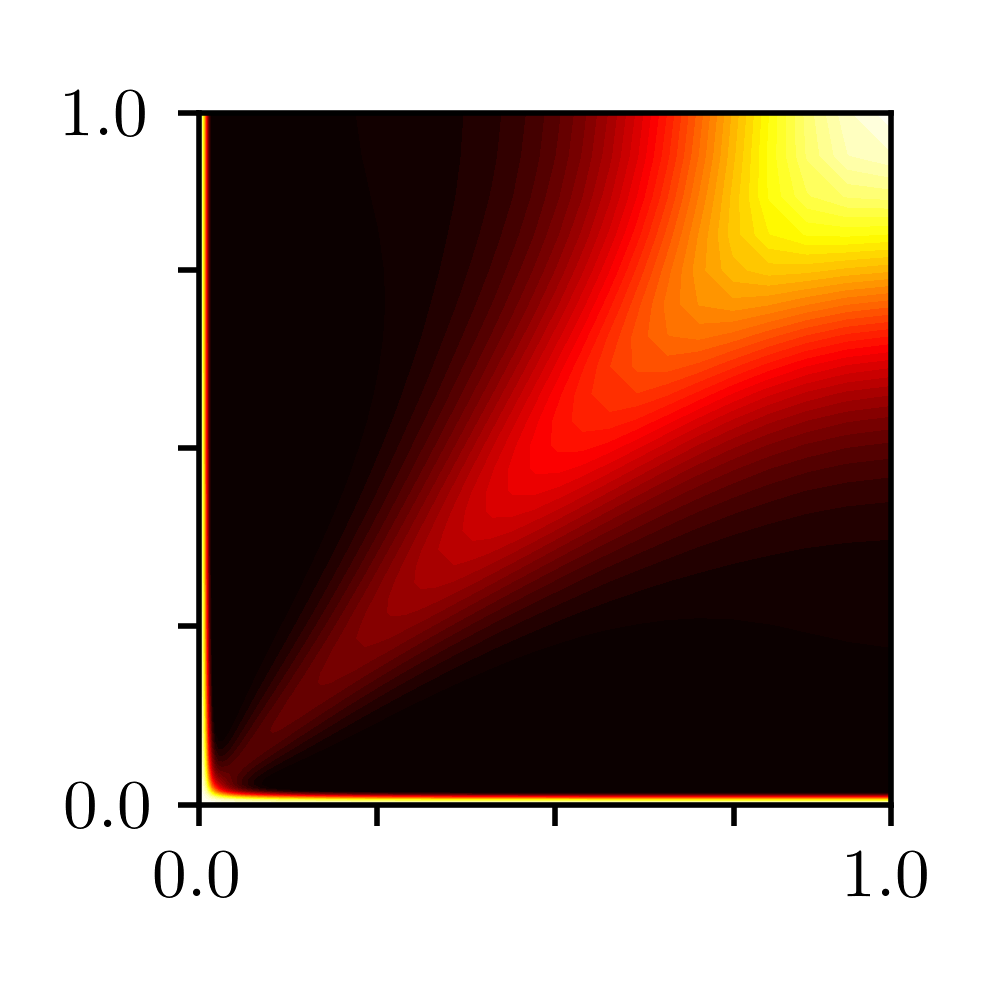}}
    \subfloat[$\| g^{(4)} \mathbf{T}^{(4)} \|$]
    {\includegraphics[width=0.2\textwidth]{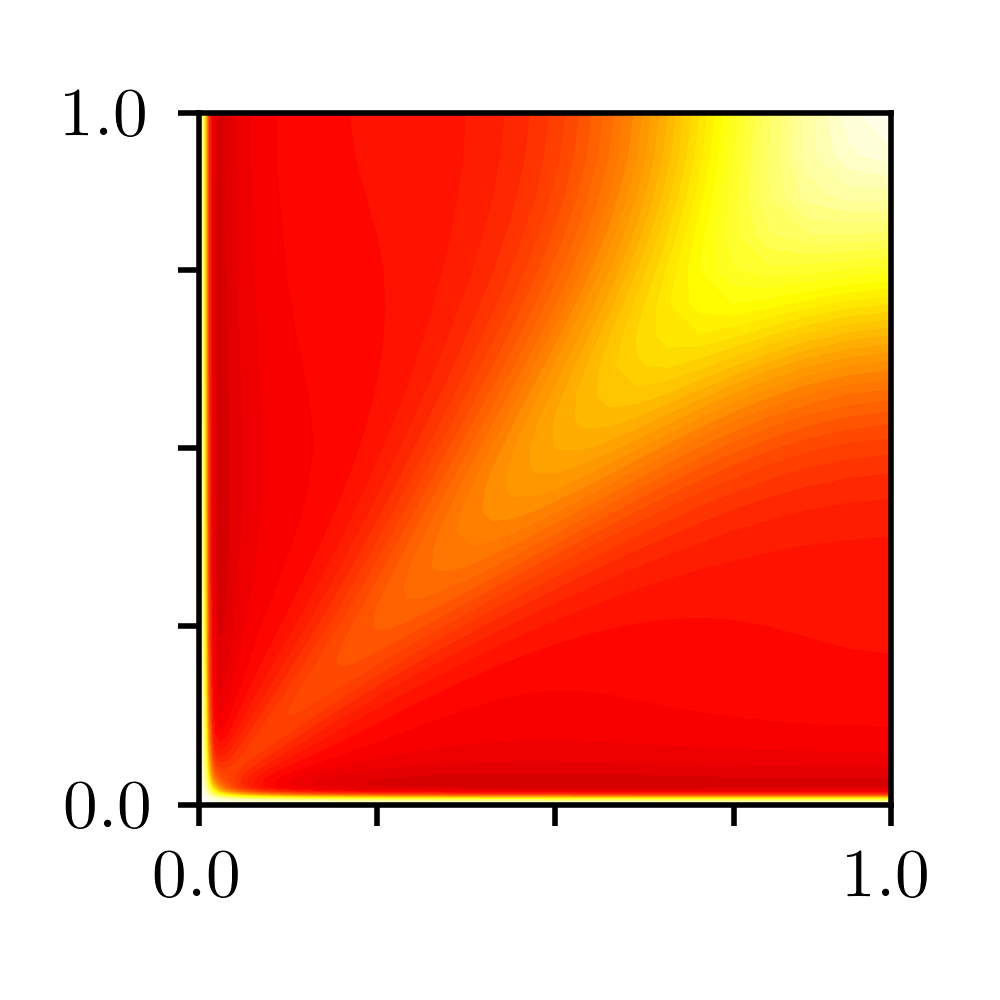}}
    \includegraphics[width=0.07\textwidth,trim=0 -0.5cm 0 0]{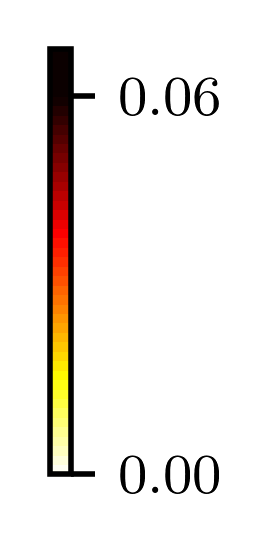}
    \caption{Learned tensor components for the square duct case}
    \label{fig:tensor_component_duct}
\end{figure}

The ensemble-based model-consistent training is flexible to provide interpretable models based on sparse observations.
It can be supported by the results of the tensor components as shown in Figure~\ref{fig:tensor_component_duct}.
It shows that the linear tensor component ~$g^{(1)}\mathbf{T}^{(1)}$ from the learned model is larger than other nonlinear components, i.e., $g^{(2)}\mathbf{T}^{(2)}$, $g^{(3)}\mathbf{T}^{(3)}$, and $g^{(4)}\mathbf{T}^{(4)}$, but at similar magnitudes.
This is in contrast to the S809 airfoil case, where the linear tensor component is larger than the nonlinear components by several orders of magnitude, as shown in Fig.~\ref{fig:tensor_component_s809}.
The relatively large magnitude of the nonlinear tensors in this case is due to the secondary flow characteristics that are driven by the imbalance of the Reynolds normal stress.
The linear tensor~$g^{(1)}\mathbf{T}^{(1)}$ cannot capture the anisotropy of the Reynolds stress, and the nonlinear components play dominant roles in predicting in-plane velocities.
Hence, for the square duct case, the ensemble-based training leads to a nonlinear model with considerable magnitude for the nonlinear terms.

In general, the ensemble-based method provides an interpretable turbulence model with appropriate nonlinearity according to limited observation data.
For instance, in the scenario of separated flows over airfoils, the optimization of linear eddy viscosity is able to remedy the deficiency on the adverse pressure gradient as shown in Figure~\ref{fig:tensor_component_s809}.
In contrast, for the secondary flow, the nonlinear terms are required to accurately estimate the imbalance of the Reynolds normal stress as shown in Figure~\ref{fig:tensor_component_duct}, which is the driving force for the spanwise vorticity.
We emphasize that the available observations are often sparse in practical applications, e.g., lift force and sparse velocity measurements, as used in this work.
Such severe ill-posedness poses challenges to the training method in learning dominant physical mechanisms with various flow characteristics.
Hence, the flexibility of ensemble-based training is demonstrated in discovering interpretable models from sparse data.

The square duct case has been used in Zhang et al. (2022) \cite{zhang_ensemble-based_2022}, which is a proof of concept for the ensemble-based learning method.
In contrast, the current study aims to demonstrate the flexibility of the ensemble method in capturing separated and secondary flows by adjusting the nonlinearity of the turbulence model.
Specifically, it is observed here that the ensemble method can learn a linear eddy viscosity model for the separated flow and a nonlinear eddy viscosity model for the secondary flow. 
This is different from the previous work~\cite{zhang_ensemble-based_2022}, which is a proof of concept for the ensemble-based learning method.
Moreover, here we use sparse DNS data to train neural network models, which shows the capability of the ensemble Kalman method to handle sparse data in realistic applications.
In contrast, in the previous work~\cite{zhang_ensemble-based_2022}, full-field flow data approximated with the quadratic model of Shih (1993) ~\cite{shih1993realizable} is used as synthetic truth, which is not identical to the DNS data.

\section{Conclusions}
\label{sec:conclusion}

This work investigates the physical interpretation of neural-network-based turbulence modeling with the ensemble Kalman method.
The observation data, including aerodynamic lift and velocity measurements, are used to train the turbulence model represented with a tensor--basis neural network.
The method is applied to the flow around the S809 airfoil and the flow in a square duct.
Both cases show that the learned model significantly improves the flow predictions, and the model improvement can be interpreted from a physical viewpoint.
In the S809 airfoil, the learned model reduces the eddy viscosity around the upstream boundary layer and captures the appropriate onset of the flow separation, which improves the prediction of the lift force compared to the baseline $k$--$\omega$ model.
The learned model can be well generalized to different angles of attack.
In the square duct case, the learned model produces a nonlinear eddy viscosity model, which captures the imbalance of the Reynolds normal stress and the in-plane velocity. 
The ensemble Kalman method can provide appropriate turbulence models based on limited observation data.
For the flow over the S809 airfoil, the training method provides an optimized linear eddy viscosity model based on the lift force measurements, which is able to capture the flow separation.
In contrast, for the flow in a square duct, the training method provides a nonlinear eddy viscosity model to estimate the anisotropy of Reynolds stress and capture the in-plane secondary flows.

\appendix

\section{Practical implementation}
\label{sec:implementation}
The practical implementation of the ensemble-based turbulence modeling framework is detailed in this subsection.
Given the observation error~$\mathsf{R}$, the data set~$\mathsf{y}$, and the sample variance~$\sigma$, the training procedure is summarized briefly below.

\begin{enumerate}
   \item Pre-training: 
    To obtain the initial weight $\bm{w}^0$ of the neural network, we pre-train the network to be equivalent to a linear eddy viscosity model such that $g^{(1)}=-0.09$ and $g^{(2-10)}=0$. The obtained weights $\bm{w}^0$ are set as the initial value for model training~\cite{strofer2021end}.
    
    \item  Initial sampling: We assume that the weights are independent and identically distributed (i.i.d.)~Gaussian random variables with mean $\bm{w}^0$ and variance $\sigma^2$.
    As such, we draw random samples of the weights through the formula $\bm{w}_j = \bm{w}^0 + \bm{\epsilon}_j$, where $\bm{\epsilon} \sim \mathcal{N}(0, \sigma^2)$. 

    \item Feature extraction: The velocity field~$\boldsymbol{u}$ and turbulence time scale~$\tau_s$ are used to compute the scalar invariants~$\bm{\theta}$ and the tensor bases~$\boldsymbol{\mathbf{T}}$  based on the equations~\eqref{eq:tensor_basis} and~\eqref{eq:scalar_invariant}. 
    The scalar invariants are normalized and then adopted as the inputs of the neural network. Further, the tensor bases are employed to construct the Reynolds stress by combining with the outputs of the neural network as illustrated in step 4.

    \item Evaluation of Reynolds stress:
    The input features~$\bm{\theta}$ are propagated to the basis coefficient $\boldsymbol{g}$ with each realization of the weights~$\bm{w}$. Then the Reynolds stress can be constructed by combining the coefficient $g$ and the tensor basis~$\boldsymbol{\mathbf{T}}$ based on Eq.~\eqref{eq:tau}.
    
    \item Propagation to mean flow fields: 
    The mean velocity is obtained by solving the RANS equations for each constructed Reynolds stress.
    Moreover, the turbulence kinetic energy and the dissipation rate are obtained by solving the turbulence transport equations.
    
    \item Update weights of neural networks: 
    The iterative ensemble Kalman method is used to update the weights of the neural network based on Eq.~\eqref{eq:enkf}.
    In the scenario of multiple observations, e.g., the S809 airfoil case in this work, the data sets are randomly shuffled and then incorporated sequentially.
    Specifically, In the S809 airfoil case, the data from two different flow conditions are  shuffled to generate a data set with random ordering.
    Then the observation data is incorporated sequentially in the shuffled order.
    The observation is reshuffled once the entire data sets are traversed.
    Besides, for each data, the Kalman update is iterated in an inner loop, and the maximum of the iteration step is set as 3 based on our sensitivity study.
    The random data ordering can escape from local minima~\cite{bottou2003stochastic} that provide good predictions for one case but inferior results for other cases based on our numerical tests.
    \end{enumerate}
    If the ensemble variance is smaller than the observation error or the total iteration maximum is reached, the training is considered converged; otherwise, continue to Step $3$ until the convergence criterion is met.

\section{Predictive performance for unseen data}
\label{sec:unseen}

We show additional prediction results of the learned model on unseen data for both the S809 airfoil case and the square duct case in this section.

\begin{figure}[!htb]
    \centering
    \includegraphics[width=0.6\textwidth]{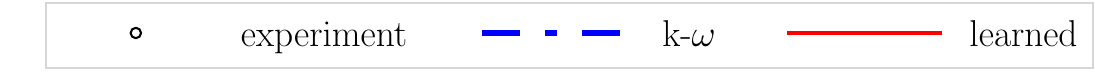} \\
    \subfloat[$\alpha=11^\circ$]{\includegraphics[width=0.4\textwidth]{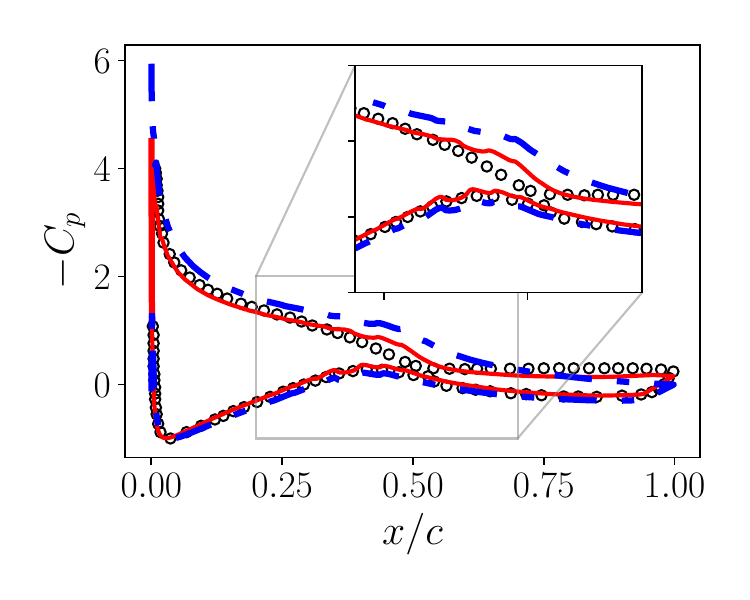}}
    \subfloat[$\alpha=18^\circ$]{\includegraphics[width=0.4\textwidth]{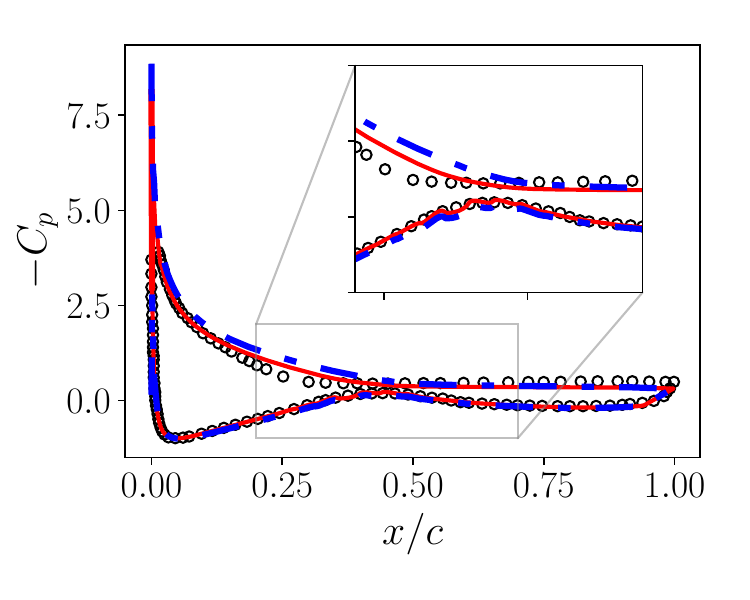}}
    \caption{Prediction of wall pressure coefficient~$C_p$ at $\alpha=11^\circ$ and $18^\circ$ with the learned model and the baseline $k$--$\omega$ model compared to the experimental data~\cite{osti_437668} for the S809 airfoil case
    }
    \label{fig:cp_test}
\end{figure}

In the S809 airfoil case, the lift force measurements from angles of attack of $8^\circ$ and $14^\circ$ are used for training, which can improve predictions of aerodynamic lift at unseen angles of attack.
Here we show that wall pressure prediction can be also improved with the learned model at unseen angles of attack.
The wall pressure predictions with the learned model in angles of attack of $11^\circ$ and $18^\circ$ are shown in Figure~\ref{fig:cp_test}, with comparison to the experimental data and prediction of the baseline $k$--$\omega$ model.
It can be seen that the baseline model underestimates the surface pressure on the suction side of the airfoil, which leads to large discrepancies in the predicted aerodynamic lift as presented in Fig.~\ref{fig:generalizability}.
In contrast, the learned model significantly improves the prediction of the wall pressure distribution and eventually the lift coefficient at both the angles, compared to the baseline $k$--$\omega$ model.

\begin{figure}[!htb]
    \centering
    \includegraphics[width=0.6\textwidth]{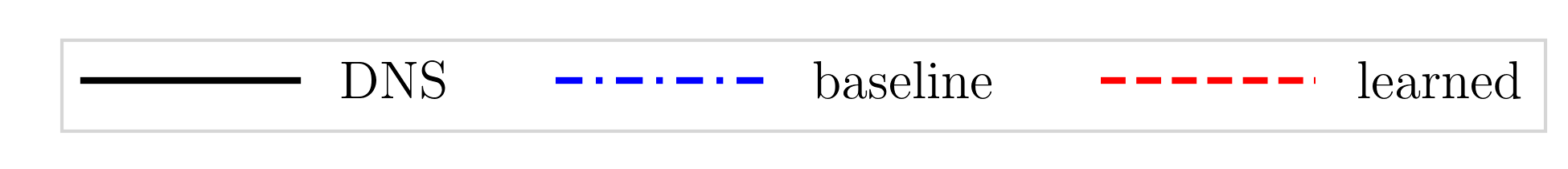} \\
    \subfloat[Velocity $u_{x}$ profiles]{\includegraphics[height=0.36\textwidth]{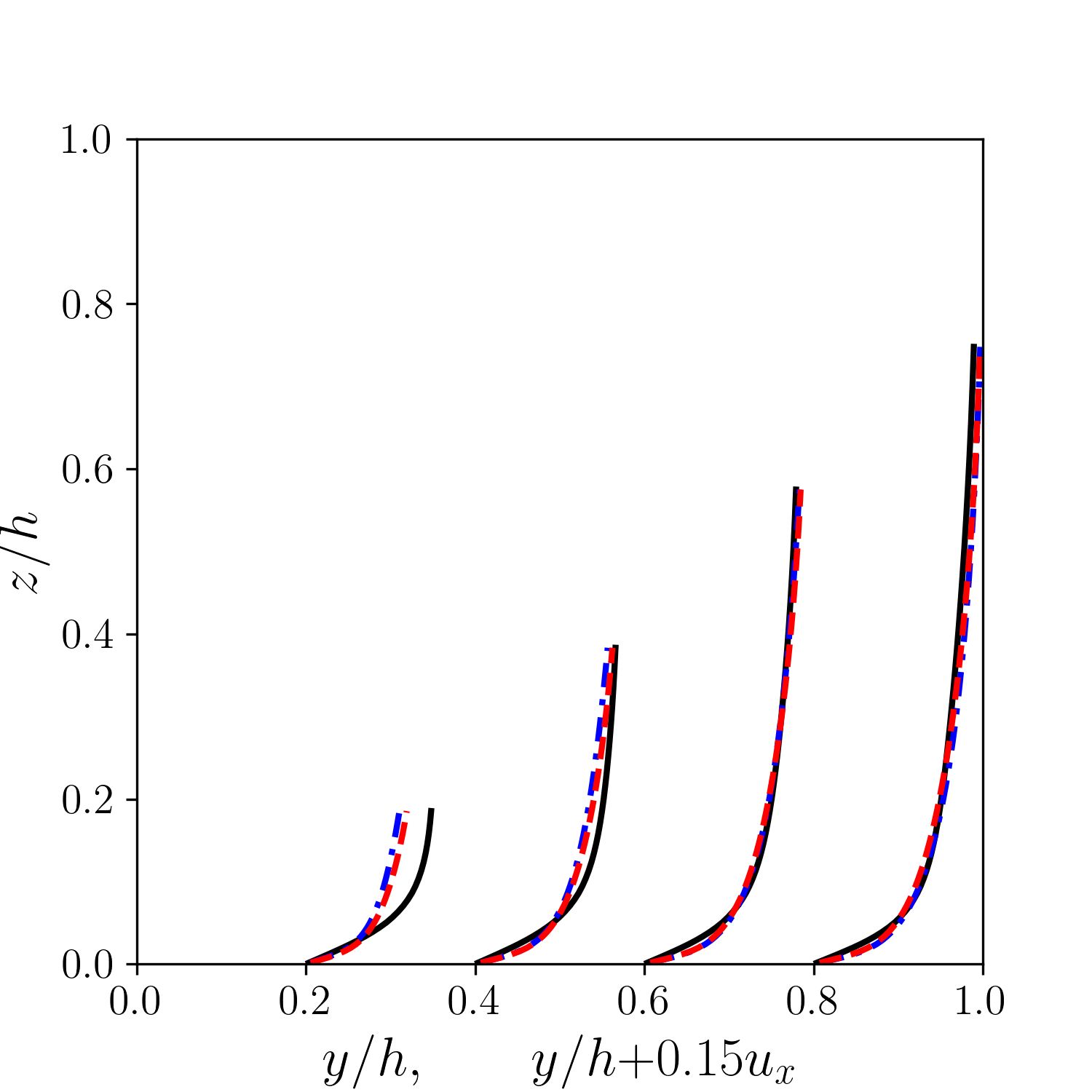}}
    \quad
    \subfloat[Velocity $u_{y}$ profiles]{\includegraphics[height=0.36\textwidth]{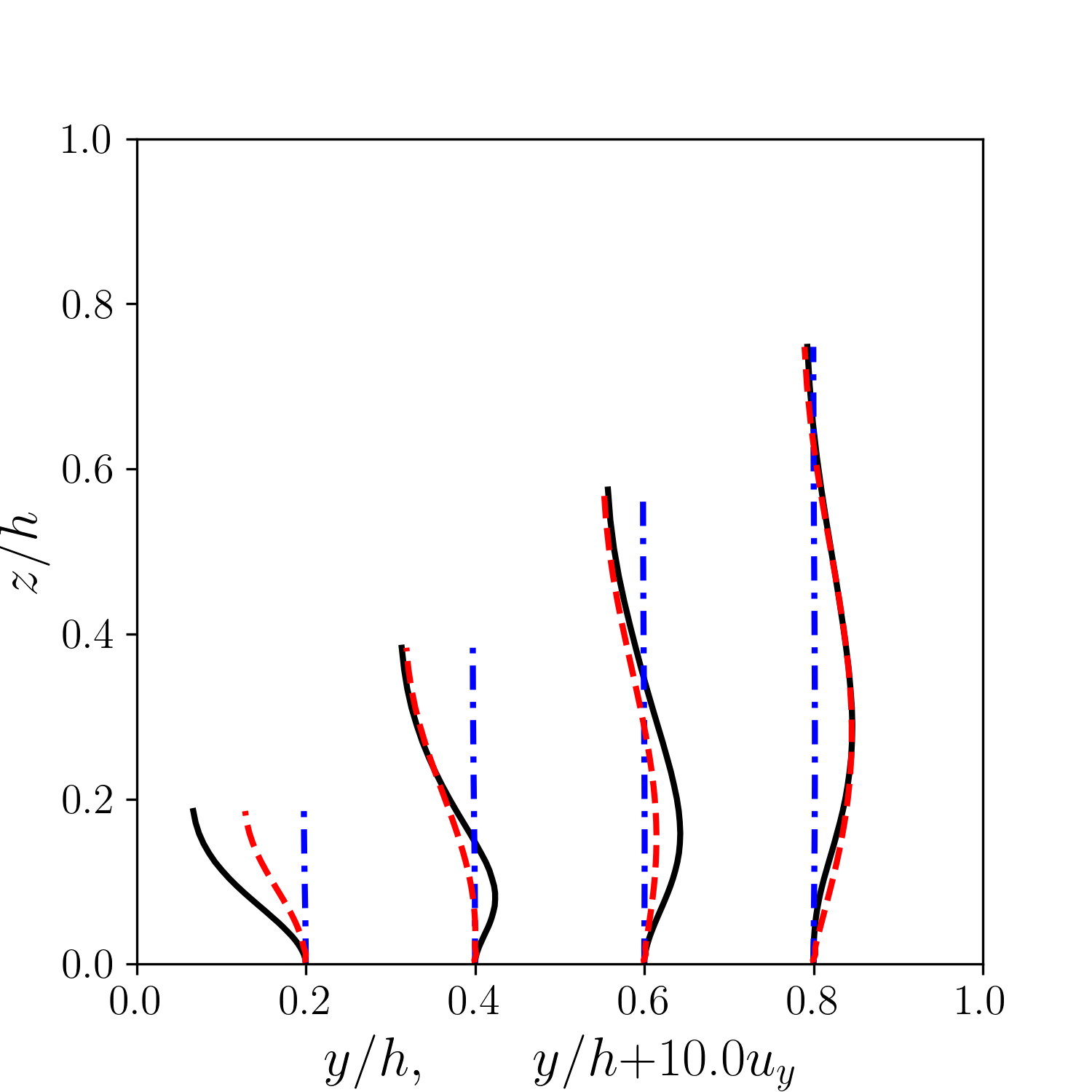}}
    \quad
    \subfloat[Reynolds Stress $\tau_{yy}-\tau_{zz}$ profiles]{\includegraphics[height=0.36\textwidth]{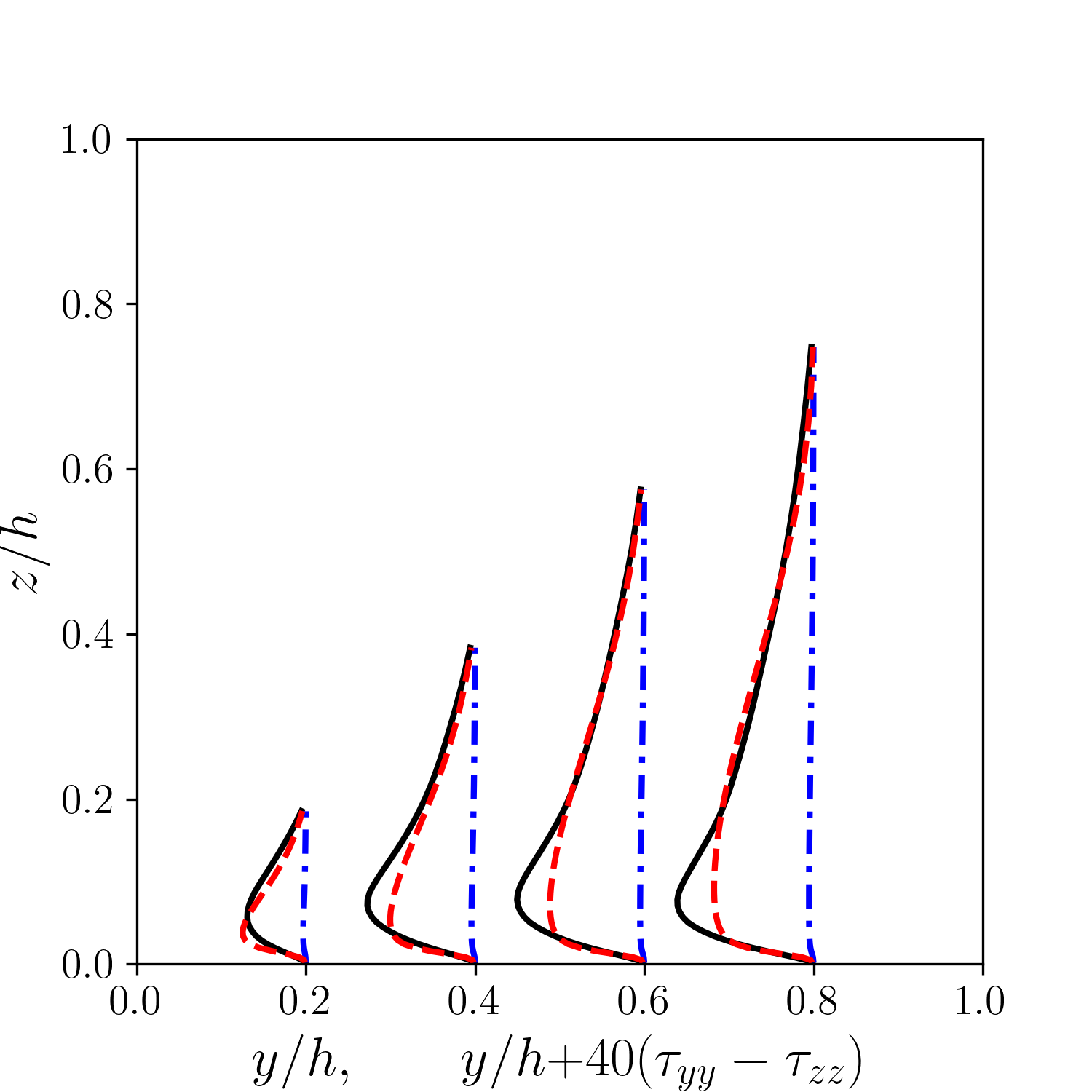}}
    \quad
    \subfloat[Reynolds Stress $\tau_{yz}$ profiles]{\includegraphics[height=0.36\textwidth]{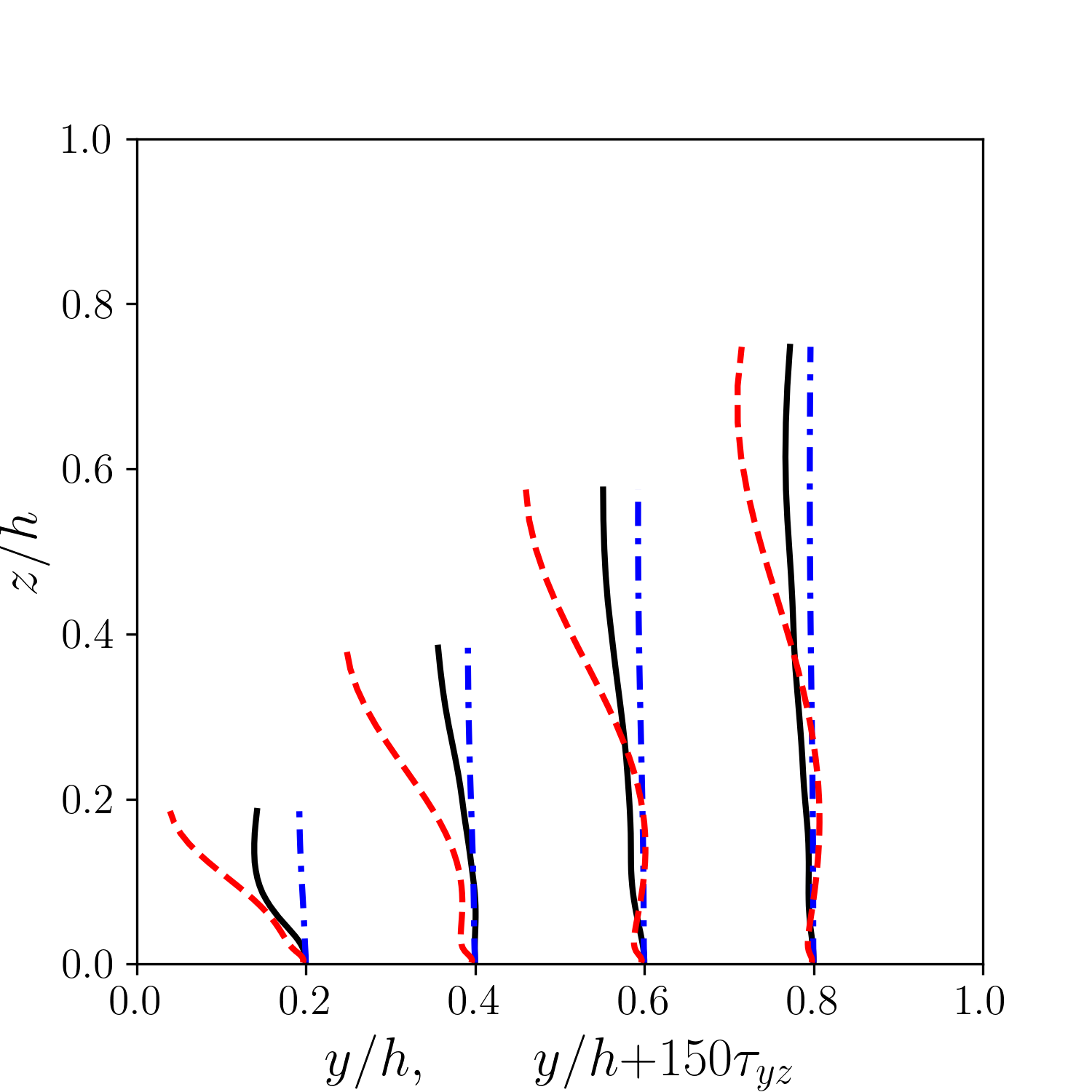}}
    \caption{Prediction of velocity and Reynolds stress along profiles at $y/H=0.2, 0.4, 0.6, 0.8$ with comparison among the learned model, the baseline model, and the experimental data, for the square duct case}
    \label{fig:sd_results_tau_profiles_add}
\end{figure}

In the square duct case, the velocity along profiles of $y/h=0.25, 0.5, 0.75,$ and $1.0$ are used for training and lead to local predictive improvement in both the velocity and the Reynolds stress.
Here we provide the model prediction at four unseen locations, i.e., $y/h=0.2, 0.4, 0.6,$ and $0.8$.
The results are shown in Fig.~\ref{fig:sd_results_tau_profiles_add}, with a comparison to the DNS data and the baseline $k$--$\varepsilon$ model.
Apparently, at these unobserved locations, the learned model predicts well the velocity component $u_y$ and the difference of the imbalance of Reynolds normal stress $\tau_{yy}-\tau_{zz}$. The learned model also yields a non-zero shear component $\tau_{yz}$, while the baseline $k$--$\varepsilon$ model yields zero shear. The latter is qualitatively incorrect based on the DNS data.

\section*{Acknowledgment}
XLZ and GH are supported by the NSFC Basic Science Center Program for ``Multiscale Problems in Nonlinear Mechanics'' (No. 11988102). XLZ also acknowledges support from the National Natural Science Foundation
of China (No.~12102435) and the China Postdoctoral Science Foundation (No.~2021M690154).  
HX acknowledges the support from the National Research Foundation of Korea (No.~NRF-2021H1D3A2A01096296) during his sabbatical visit to Gwangju Institute of Science and Technology, where this work was performed.

\end{document}